\documentclass[aps,pre,reprint,10pt,superscriptaddress,showpacs]{revtex4-1}
\usepackage{amsfonts} 
\usepackage{amsmath}
\usepackage{amssymb}
\usepackage{graphicx}
\usepackage{subfigure}
\usepackage{color}
\newcommand*\xbar[1]{%
  \hbox{%
    \vbox{%
      \hrule height 0.5pt 
      \kern0.5ex
      \hbox{%
        \kern-0.1em
        \ensuremath{#1}%
        \kern-0.1em
      }%
    }%
  }%
} 
\begin{document}
\title{Landau damping of Gardner solitons in  a  dusty bi-ion  plasma }
\author{A. P. Misra}
\email{apmisra@visva-bharati.ac.in; apmisra@gmail.com}
\author{Arnab Barman}
\affiliation{Department of Mathematics, Siksha Bhavana, Visva-Bharati University, Santiniketan-731 235, West Bengal, India}
\begin{abstract}
 The effects of linear Landau damping on the nonlinear propagation of dust-acoustic solitary waves (DASWs) are studied  in a collisionless unmagnetized  dusty   plasma with two species of positive ions.  The  extremely massive, micron-seized, cold and negatively charged dust particles are described by fluid equations, whereas  the two species of positive ions, namely the cold (heavy) and hot (light)  ions are described by the kinetic Vlasov equations.   Following   Ott and Sudan [Phys. Fluids {\bf 12}, 2388 (1969)], and by considering lower and higher-order perturbations,  the evolution of DASWs with Landau damping is shown to be governed by   Korteweg-de Vries (KdV), modified KdV (mKdV) or Gardner  (KdV-mKdV)-like equations.  The properties of the phase velocity and the Landau damping rate of DASWs are studied for different values of the ratios of the temperatures $(\sigma)$ and the number densities $(\mu)$   of hot and cold ions as well the cold to hot ion mass ratio $m$. The distinctive features of the  decay rates of the  amplitudes of the KdV, mKdV and Gardner  solitons   with a small effect of Landau damping are also studied in different parameter regimes. It is found that the Gardner soliton points to lower wave amplitudes than the KdV and mKdV solitons.   The results may be useful for understanding the localization of solitary pulses and associated wave damping (collisionless) in laboratory and space plasmas (e.g., the F-ring of Saturn)  in which the number density of free electrons is much smaller than that of ions and the heavy, micron seized dust grains are highly charged.  
\end{abstract}
\received{31 March 2015}
\revised{10 June 2015}
\maketitle
\section{Introduction}
Dusty plasmas with an admixture of two species of ions are ubiquitous in  both   space (e.g., Earth's D and lower E regions, Titan's atmosphere, F-ring of Saturn etc.) \cite{narcisi1971,coates2007} and  laboratory plasmas \cite{kim2006,kim2013}. In addition to the main positive ion (light or hot),   dusty plasmas also contain   a second,  micron-sized positive or negative ion (heavy or cold), which  characterizes the appearance of an additional wave mode  in the plasma \cite{kim2006,kim2013}. There are in fact several motivations for investigating collective plasma oscillations and nonlinear coherent structures like solitons and double layers in dusty plasmas with two groups of ions (See, e.g., Refs. \onlinecite{kim2006,kim2013,rosenberg2007,misra2012,misra2013,hatami2009,mamun1999}). In planetary rings, in particular, in the F-ring of Saturn an anomalous situation may arise in which massive dust grains capture almost all the electrons from the background plasma, i.e., the number density of free electrons becomes much smaller than the ions \cite{goertz1989}.  Such density depletion of electrons (by aerosol particles)  have also been observed in the summer polar mesosphere at about $85$ km altitude \cite{reid1990} as well as in the laboratory \cite{barkan1995}. On the other hand,  plasmas with two species of positive ions have been found to have applications in secondary electron emission guns (e.g., used in bacterial inactivation) in which helium   and argon are used as the main gas and an auxiliary gas  for ionizing helium atoms \cite{chalise2004}.

In dusty plasmas,  the size of the dust grains  is, in general, not a constant but  may vary in the range of $0.05-10\mu$m, their mass $m_d\sim\left(10^6-10^{12}\right)m_i$, where $m_i$ is the ion mass, and they have atomic numbers $z_d\sim10^3-10^5$ (See,e.g., Ref. \onlinecite{shukla2002}). It is well known that the  presence of charged dust grains modifies the dispersion properties of  plasma waves   and   gives  rise to a new low-frequency wave mode, called dust-acoustic wave (DAW), which was first theoretically predicted by Rao \textit{et al.} \cite{rao1990}. These waves, associated with the motion of highly charged and massive dust grains, occur at relatively long time ($\sim0.1$ s) and space scales ($\sim1$ cm). Furthermore, in dusty plasmas, the presence of second ion species  can significantly modify not only the dispersion properties of DAWs but also some nonlinear coherent structures (See, e.g., Refs. \onlinecite{misra2012,misra2013,hatami2009,mamun1999}). These waves can admit collisionless damping due to the resonance interaction of trapped and/or free particles  with the wave, i.e., when the particle's velocity becomes closer to the wave phase velocity  \cite{landau1946}. Such wave damping was first theoretically predicted by Landau \cite{landau1946} in $1946$, and later confirmed experimentally by Malmberg \textit{et al.} \cite{malmberg1964}. Motivated by this invention,  Ott and Sudan \cite{ott1969,ott1970} first studied  the effects of (linear) electron Landau damping on the nonlinear propagation of  ion-acoustic solitons  in electron-ion plasmas. They had  neglected the particle's trapping effects  on the assumption that the particle trapping time is much longer than that of Landau damping. Their evolution equation was  in the form of a Korteweg-de Vries (KdV) equation which contains, apart from the lowest order nonlinear terms and the term that corresponds to linear dispersion relation of ion-acoustic waves,  a source term that models the lowest-order effects of resonant particles. In their work, they showed that  a wave steepening of an initial wave form may occur  or not depending on the relative size of the nonlinearity compared to the Landau damping. The latter  was also shown to slow down the wave amplitude with time. 

Recently,  a number of attempts have been made to study the effects of Landau damping on   nonlinear  solitary waves owing to its importance in different plasma environments. To mention few, Mukherjee \textit{et al.} \cite{mukherjee2014} studied the effects of linear Landau damping on the nonlinear ion-acoustic solitons in a quantum plasma with the description of KdV equation. The similar investigation was made by Barman {\it et al.} \cite{barman2014} for nonlinear dust-acoustic solitary waves   in a dusty pair-ion plasma.   In a linear theory, the effects of dust size distribution on the Landau damping rates have also been studied by Zhang {\it et al.} in a multi-component dusty plasma \cite{zhang2014}.   However, all the above works were based on either KdV equation for nonlinear waves or linear dispersion laws for damping rates. As is well known that the KdV solitons are formed for parameter values well below the critical values (at which the nonlinear coefficient of the KdV equation vanishes) and the modified KdV (mKdV) solitons are formed for parameter values close to the  critical values. However, when both the KdV and mKdV equations fail,   Gardner  (or KdV-mKdV) equation, which generalizes the KdV and mKdV equations, is more appropriate to describe the evolution of solitary waves in plasmas. Such (Gardner) solitons are formed for parameters very close to and around the critical values of the plasma parameters.    Though, there are a number of works in the literature dealing with the Gardner solitons in plasmas (See, e.g., Ref. \onlinecite{mamun2014}), however, all are based on fluid theory approach and thus without  the effects of  Landau damping.   

In this work,  we    investigate  the Landau damping effects on dust-acoustic (DA) solitary waves (SWs) (DASWs) in a dusty   plasmas with   two different species of positive ions namely heavy or cold and light or hot ions.  Following Ott and Sudan \cite{ott1969,ott1970}, we derive a KdV as well as mKdV and Gardner equations (with Landau damping) by considering lower and higher order nonlinearities to describe  the evolution of dust-acoustic solitary waves for parameter values far, near and around their critical values. The properties of the KdV, mKdV and  Gardner solitons with the effects of Landau damping, the linear phase velocity as well as the Landau damping rate are  also analyzed with the effects of the ratios of the number densities $(\mu)$  and temperatures $(\sigma)$   of hot to cold ions as well as cold to hot ion mass ratio $m$. It is shown that the landau damping has an effect  of slowing down the wave amplitude with time, and the Gardner soliton points to lower wave amplitude.   
\section{Basic Equations}      
     We consider the nonlinear propagation of dust-acoustic waves (DAWs) in an unmagnetized collisionless electron free dusty plasma which consists of extremely massive, negatively charged, micron-seized cold dust fluids and singly charged hot (light) and cold (heavy)  positive ions. We assume that the  dust particles have constant   mass and charge. The latter, in general, can vary and   introduce  a new low-frequency wave eigenmode as well as  a  dissipative effect (other than the wave damping) into the system. However, we   neglect  this charge fluctuation effect   on the assumption that the charging rate of dust grains is very high compared to the dust plasma oscillation frequency.    We also assume that the size of the dust grains is small compared to the average interparticle distance, and  the ratio   of electric charge to mass of the dust grains remains much smaller than those of ions.      At equilibrium, the overall charge neutrality condition satisfies   
    \begin{equation}
    n_{c0}+n_{h0}=z_dn_{d0}, \label{charge-neutrality}
    \end{equation}
  where $n_{j0}$ is the unperturbed number density of species $j$ (where $j$=$c,~h$ and $d$, respectively, stand for cold ions, hot ions  and   charged dust grains) and $z_d$ ($>0$) is the unperturbed  dust charge state.

The basic equations for the dynamics of charged dust particles together with cold and hot ions in one space dimension are 
   \begin{eqnarray}
   \partial_t n_d+\partial_x (n_du_d)=0, \label{cont-eqn}
   \end{eqnarray}
   \begin{eqnarray}
   \partial_t u_d+u_d\partial_x u_d=-({q_d}/{m_d})\partial_x \phi, \label{montm-eqn}
   \end{eqnarray}
   \begin{eqnarray}
   \partial_t f_j+v\partial_x f_j-({q_j}/{m_j})\partial_x \phi\partial_v f_j=0, \label{Vlasov-eqn}
   \end{eqnarray}
   \begin{eqnarray}
   \partial^2_x\phi=-4\pi e(n_c+n_h-z_dn_d), \label{Poisson-eqn}
   \end{eqnarray}
  where $\partial_\alpha f$ denotes partial differentiation of $f$ with respect to $\alpha$. Also,  Eq. \eqref{Vlasov-eqn} is for the cold $(j=c)$ and hot $(j=h)$ ions with their number densities  given by 
 \begin{eqnarray}
 n_j=\int_{-\infty}^{\infty} f_j dv. \label{density-eqn}
 \end{eqnarray}
  In equations \eqref{cont-eqn}-\eqref{density-eqn}, the symbols $n_d$, $u_d$, $q_d$ ($=-z_de,~ e$ being the elementary charge), $m_d$, respectively, denote the number density, fluid velocity, charge and mass of dust grains. Also, $v$ is the particle's velocity,   $f_j$ is the   velocity distribution function and $m_j$  is the  mass   of hot  and cold  ions. Furthermore, $q_j=e$,  $\phi$ is the electrostatic potential, and $x$ and $t$ are the space and time coordinates. 

In what follows, we normalize the physical quantities according to $u_d\rightarrow u_d/c_s$, $\phi\rightarrow e\phi/k_BT_h$, $n_j\rightarrow n_j/n_{j0}$, $n_d\rightarrow n_d/n_{d0}$, $f_j\rightarrow f_jv_{tj}/n_{j0}$, $v\rightarrow v/v_{th}$, where $c_s=\sqrt {z_dk_BT_h/m_d}=\omega_{pd}\lambda_D$ is the DA speed with $\omega_{pd}=\sqrt{4\pi n_{d0}z^2_d e^2/m_d}$ and $\lambda_D=\sqrt{k_BT_h/4\pi n_{d0}z_d e^2}$ denoting, respectively, the dust plasma frequency and the plasma Debye length. Here, $k_B$ is the Boltzmann constant, $T_j$ is the thermodynamic temperature   and $v_{tj}(=\sqrt{k_BT_j/m_j})$ is the thermal velocity of  $j$-species particles. The space and time variables are normalized by $L$ and $L/c_s$ respectively, where $L$ is the characteristic scale length for variations of the dependent variables $n_j,~u_d,~\phi,~f_j$ etc.   
   Thus, from Eqs. \eqref{cont-eqn}-\eqref{density-eqn}, we obtain the following set of equations in dimensionless form:  
   \begin{eqnarray}
   \partial_t n_d+\partial_x (n_du_d)=0, \label{cont-eqn-nond}
   \end{eqnarray}
   \begin{eqnarray}
   \partial_t u_d+u_d\partial_x u_d=\partial_x \phi,\label{montm-eqn-nond}
   \end{eqnarray}
   \begin{eqnarray}
   \delta\partial_t f_j+v\partial_x f_j-({m_h}/{m_j})\partial_x \phi\partial_v f_j=0, \label{Vlasov-eqn-nond}
   \end{eqnarray}
   \begin{eqnarray}
   \left(\lambda_D^2/L^2\right)\partial^2_x\phi=-(\mu_{cd}n_c+\mu{hd}n_h+n_d), \label{Poisson-eqn-nond}
   \end{eqnarray}
   \begin{eqnarray}
   n_j=\sqrt{m_j T_h/ m_h T_j}\int_{-\infty}^{\infty} f_j dv,\label{density-eqn-nond}
   \end{eqnarray}
   where $\delta=\sqrt{z_dm_h/m_d}$ and  $\mu_{jd}=n_{j0}/z_dn_{d0}$ is the density ratio for $j=c,~h$.
 The charge neutrality condition \eqref{charge-neutrality} reduces to 
   \begin{equation}
   \mu_{cd}+\mu_{hd}=1,~\text{or}~ \mu_{hd}=\mu/(1+\mu), \label{charge-neutrality-nond}
   \end{equation}
   where $\mu=n_{h0}/n_{c0}$ is the ratio of hot to cold unperturbed ion number densities.
 As in Ref. \onlinecite{barman2014}, there are three basic parameters. The first one is the ratio $\delta\equiv\sqrt{z_dm_h/m_d}$ or $m\delta$ ($m=m_c/m_h$), which represents the finite inertial effects of  hot and cold ions, and, in particular, the Landau damping. The second one is the ratio of perturbed dust density to its equilibrium value $n_{d1}/n_{d0}$, which measures the strength of  nonlinearity in electrostatic perturbations.  Lastly, the parameter $\lambda_D/L$, measuring the strength of the wave dispersion due to deviation from the quasineutrality.  This parameter will eventually disappear in the left-hand side of Eq. \eqref{Poisson-eqn-nond}, if one considers the normalization of $x$ by $\lambda_D$  instead of  $L$.  Note that since the dust particles are assumed to be  cold,   the Landau damping effects are provided solely by the hot and cold ions, and the damping rate is $\propto$   $\delta$. 
 \section{Derivation of KdV equation} Before going to derive the KdV equation   we first consider the following scaling for the basic parameters described in the previous section as   \cite{ott1969,barman2014} 
\begin{equation}
 \delta=\alpha_1\epsilon,~~n_{d1}/n_{d0}=\alpha_2\epsilon,~~\lambda_D^2/L^2=\alpha_3\epsilon, \label{scale} 
\end{equation}
 where $\epsilon~(\lesssim1)$ is a smallness positive parameter and $\alpha_j\sim o(1)$ ($j=1,2,3$) is a constant.   Under the scaling \eqref{scale}    Eq. \eqref{Vlasov-eqn-nond}    for hot and cold  ions  reduces to
 \begin{eqnarray}
 \alpha_1\epsilon\partial_t f_h+v\partial_x f_h-\partial_x\phi\partial_v f_h=0,\label{p_Vlasov-eqn-nond}
 \end{eqnarray}
 and 
  \begin{eqnarray}
 \alpha_1\epsilon\partial_t f_c+v\partial_x f_c-(1/m)\partial_x\phi\partial_v f_c=0.\label{n_Vlasov-eqn-nond}
 \end{eqnarray}
 
Next, in  the small-amplitude limit (i.e., $\epsilon\rightarrow0$) in which inertial effects of  hot and cold ions are neglected, and for $L\gg\lambda_D$ (valid for long-wavelength perturbations), Eqs. \eqref{cont-eqn-nond}-\eqref{Poisson-eqn-nond} yield the following linear dispersion law \cite{barman2014}
 \begin{equation}
 v_p\equiv\omega/k=\left(\mu_{hd}+\sigma\mu_{cd}\right)^{-1/2}, \label{disp-relation}
 \end{equation}
where  $\omega$ ($k$) is the wave frequency (number) of the plane wave perturbations and $\sigma=T_h/T_i$ is the ratio of  hot  to cold ion temperatures. Equation \eqref{disp-relation} has  the  same form as Eq. (15)  in Ref. \onlinecite{barman2014} obtained in dusty plasmas with positive  and negative ions. From Eq. \eqref{disp-relation} it is seen that  the  DAWs become dispersionless in the long-wavelength oscillations with the constant phase speed $v_p~(<c_s)$. This implies that the time derivatives of all physical quantities should vanish in a frame moving at the speed $v_p$, and one can thus   expect slow variations of the wave amplitude in it. So, we introduce  the stretched coordinates   as \cite{taniuti1969}
 \begin{equation}
\xi=\epsilon^{1/2} (x-Mt),~\tau=\epsilon^{3/2}t, \label{stretching}
 \end{equation}
where   $M$ is the nonlinear wave speed (relative to the frame with $x$ and $t$ as coordinates)  normalized by $c_s$, to be shown  to be equal to $v_p$ later.  
 The dynamical variables are expanded   as
\begin{eqnarray}
n_d&&=1+\alpha_2\epsilon n^{(1)}_d+\alpha^2_2\epsilon^2 n^{(2)}_d+\cdots,\notag \\
u_d&&=\alpha_2\epsilon u^{(1)}_d+\alpha^2_2\epsilon^2 u^{(2)}_d+\cdots,\notag \\
\phi&&=\alpha_2\epsilon\phi^{(1)}+\alpha^2_2\epsilon^2\phi^{(2)}+\cdots,\label{expansions}\\
n_j&&=1+\alpha_2\epsilon n^{(1)}_j+\alpha^2_2\epsilon^2 n^{(2)}_j+\cdots,\notag \\
f_j&&=f^{(0)}_j+\alpha_2\epsilon f^{(1)}_j+\alpha^2_2\epsilon^2f^{(2)}_j+\cdots,\notag
\end{eqnarray}
where the equilibrium distribution function $f^{(0)}_j$,   $j=c,~h$, are assumed to be the Maxwellian 
\begin{equation}
f^{(0)}_j=\sqrt{{1}/{2\pi}}\exp\left[(m_jT_e/m_eT_j)(-v^2/2)\right].\label{f_0 Maxwellian distribution}
\end{equation}

\begin{figure}[!htb]
\centering 
{\includegraphics[height=2.0in,width=3.5in]{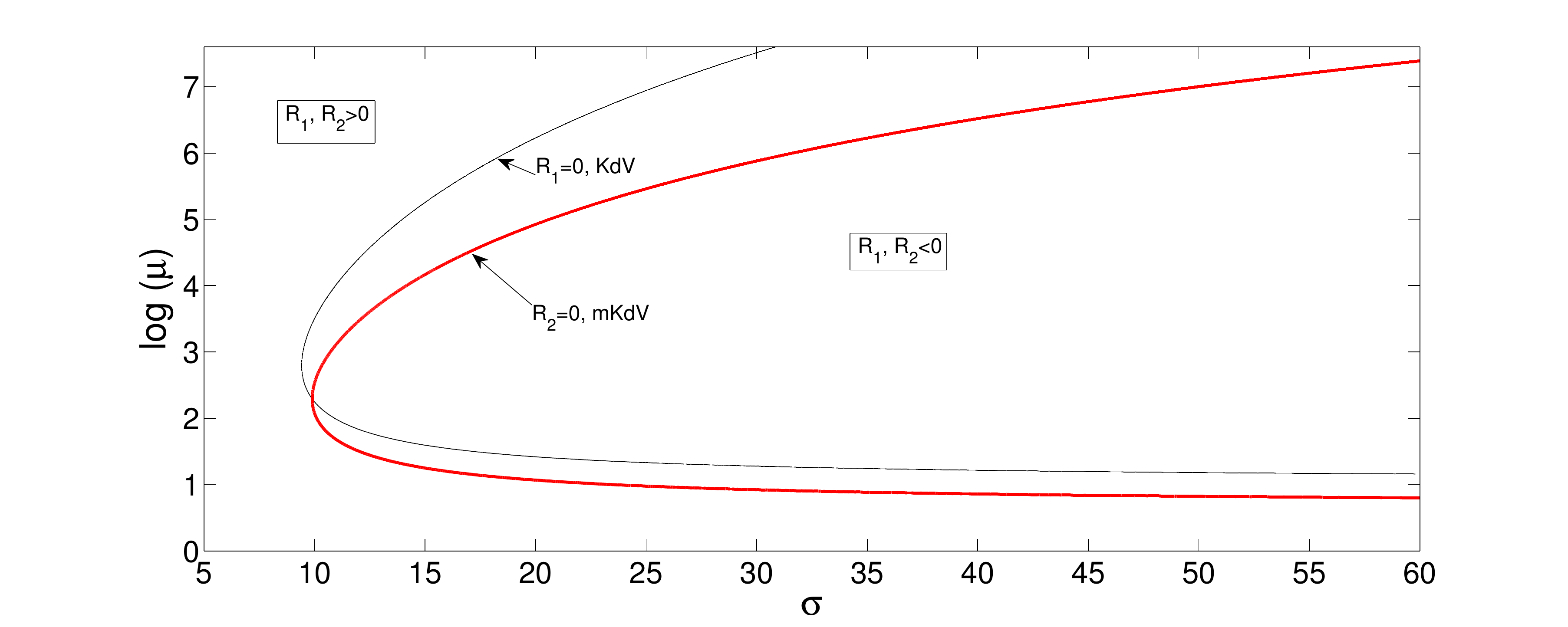}}
{\caption{Contour plots of $R_1\equiv 3-\sigma_2/\sigma_1^2\equiv3-(1+\mu)(\sigma^2+\mu)/(\sigma+\mu)^2=0$ (The thin or black curve) [i.e., vanishing of the nonlinear coefficients $b$ and $b_1$ of the KdV and Gardner equations \eqref{K-dV} and \eqref{mK-dV}]  and $R_2\equiv 1-\sigma_3/15\sigma_1^3\equiv1-(1+\mu)^2(\sigma^3+\mu)/(\sigma+\mu)^3=0$ (The thick or red curve) [i.e., vanishing of the nonlinear coefficient   $b_2$ of the mKdV equation  \eqref{mkdv}]  showing the   critical regions in the $\sigma-\mu$ plane. There is a common critical point $(\sigma,\mu)\approx(10,10)$ at which both the KdV and mKdV equations fail to describe the evolution of DAWs.   }
\label{fig:fig1}}
\end{figure}

In the next step we substitute the expressions from Eqs. \eqref{stretching} and \eqref{expansions}  into Eqs. \eqref{cont-eqn-nond}, \eqref{montm-eqn-nond}, \eqref{Poisson-eqn-nond}, \eqref{density-eqn-nond}, \eqref{p_Vlasov-eqn-nond} and \eqref{n_Vlasov-eqn-nond}, and equate different powers of $\epsilon$. In this way, we obtain the equations for   first and second order perturbations. Since the method and intermediate steps  for the derivation of the KdV equation are almost similar to that in Ref. \onlinecite{barman2014},      we skip presenting first and second order  expressions for the perturbations. However, those can be given for the derivation of the Gardner equation. The mKdV equation will then directly follow from the Gardner equation. Thus, following  Ref. \onlinecite{barman2014},  we obtain the following KdV equation
\begin{equation}
\frac{\partial n}{\partial\tau}+a\text{P}\int^{\infty}_{-\infty}\frac{\partial n}{\partial\xi'}\frac{d\xi'}{\xi-\xi'}+bn\frac{\partial n}{\partial\xi}+c\frac{\partial^3 n}{\partial\xi^3}=0,\label{K-dV}
\end{equation}
where  the coefficients of the Landau damping $(a)$, the nonlinear $(b)$ and the dispersive term $(c)$, respectively, are given by
\begin{eqnarray}
a=\left(\alpha_1/\sqrt{8\pi\sigma}\sigma_1^2\right)\left[\sqrt{m}-(\sqrt{m}-\sigma^{-3/2})\mu_{hd}\right],\label{a}
\end{eqnarray}
\begin{equation}
b=\left(\alpha_2/2\sqrt{\sigma\sigma_1}\right)\left(3-\sigma_2/\sigma_1^2\right). \label{nonl-kdv}
\end{equation}
\begin{equation}
c=(\alpha_3/2)(\sigma\sigma_1)^{-3/2},\label{c}
\end{equation}
with $\sigma_1=1+(\sigma^{-1}-1)\mu_{hd}\equiv\mu_{cd}(\sigma+\mu)/\sigma$, $\sigma_2=1+(\sigma^{-2}-1)\mu_{hd}\equiv\mu_{cd}(\sigma^2+\mu)/\sigma^2$ and
$\sigma_3=1+(\sigma^{-3}-1)\mu_{hd}\equiv\mu_{cd}(\sigma^3+\mu)/\sigma^3$.   
We note that  the nonlinear coefficient $b$ of the KdV equation \eqref{K-dV} vanishes for $R_1\equiv 3-\sigma_2/\sigma_1^2\equiv3-(1+\mu)(\sigma^2+\mu)/(\sigma+\mu)^2=0$ along a curve in the $\sigma-\mu$ plane (See Fig. \ref{fig:fig1}).  In this case, the KdV equation fails to describe the nonlinear evolution of DASWs, and one has to look for some higher order perturbations, i.e., the evolution of DASWs can then be described by mKdV or Gardner equations to be obtained shortly.
Furthermore, in absence of the Landau damping the KdV equation admits compressive or rarefactive soliton solutions according to when  $b>0~(b<0)$, i.e.,  when $b$ falls in the region $R_1>0~(R_1<0)$. We will, however, later see that the region for which $b<0$ is inadmissible for the existence of KdV solitons with a small effect of the Landau damping [See Eq. \eqref{tau0-kdv} in Sec. VI]. We also  mention that the KdV solitons with Landau damping are formed for values of the parameters $\sigma$ and $\mu$ far below  the critical values along the curve $R_1=0$, the mKdv solitons are formed for the parameters   close to the critical values and Gardner solitons are formed approximately at and around the critical values of    $\sigma$ and $\mu$.  Thus, the KdV equation may be valid for the parameters $\sigma$ and $\mu$  satisfying $R_1\sim o(1)$. However, in the case of $R_1\sim o(\epsilon)$ or $R_1\approx 0$, one has to deal with mKdV or Gardner equations which involve higher-order (of $\epsilon$) perturbations for the evolution of dust-acoustic solitary waves in the plasma. 
 
 Figure \ref{fig:fig1} shows the contour plot of $R_1=0$ (the thin or black curve) in the $\sigma-\mu$ plane in which the nonlinear coefficients  $b$ and $b_1$, respectively, of   the KdV   and Gardner (obtained later)  equations \eqref{K-dV} and \eqref{mK-dV} vanish. The plot of $R_2=0$ (The thick or red curve) represents the same, however, for the vanishing of the nonlinear coefficient $b_2$ of the mKdV equation \eqref{mkdv} (shown later).  We find that when  $\sigma$ and $\mu$ assume values satisfying $R_1\approx0$ the KdV solitons may not exist, the evolution of dust-acoustic solitons can then be described by the   mKdV equation. The latter can also be inappropriate for the evolution of solitons  when   $R_2\approx0$ for some values of $\sigma$ and $\mu$. From Fig. \ref{fig:fig1} it is also seen that the curves for $R_1=0$ and $R_2=0$ has a common point of intersection  $(\sigma,~\mu)\approx(10,10)$ at which  both the KdV and mKdV  solitons cease to exist. Also, the lower parts of these curves   below the intersection point are close to each other, implying that the parameters close to the lower part of $R_1=0$ may be some what close to that of $R_2=0$. In this case, the Gardner equation is necessary for the evolution of DASWs, which will be shown later.  
 \section{Derivation of Gardner and mKdV equations} 
To derive the Gardner equation  for DASWs we consider  $R_1\sim o(\epsilon)$. In this case,  the scaling for the basic parameters [Eq. \eqref{scale}] and the equations for hot and cold ions, i.e., Eqs. \eqref{p_Vlasov-eqn-nond} and \eqref{n_Vlasov-eqn-nond} are to be changed by the substitution $\epsilon\rightarrow\epsilon^2$ as we are looking for perturbations with higher-order (of $\epsilon$) effects. Thus, we have the modified scaling for the basic parameters
\begin{equation}
 \delta=\alpha_1\epsilon^2,~~n_{d1}/n_{d0}=\alpha_2\epsilon^2,~~\lambda_D^2/L^2=\alpha_3\epsilon^2. \label{scale-gardner} 
\end{equation}
The modified  kinetic equations for ions are  
 \begin{eqnarray}
 \alpha_1\epsilon^2\partial_t f_h+v\partial_x f_h-\partial_x\phi\partial_v f_h=0,\label{p_Vlasov-eqn-nond-gardner}
 \end{eqnarray}
 and 
  \begin{eqnarray}
 \alpha_1\epsilon^2\partial_t f_c+v\partial_x f_c-(1/m)\partial_x\phi\partial_v f_c=0.\label{n_Vlasov-eqn-nond-gardner}
 \end{eqnarray}
Also, the stretched coordinates  are redefined as 
\begin{equation}
\xi=\epsilon (x-Mt),~\tau=\epsilon^{3}t, \label{stretch-gardner}
 \end{equation}
However, the dynamical variables are   expanded in the same way as Eq. \eqref{expansions} for the KdV equation.
Next,  we substitute the expressions from Eqs.  \eqref{expansions} and \eqref{stretch-gardner}  into Eqs. \eqref{cont-eqn-nond}, \eqref{montm-eqn-nond}, \eqref{Poisson-eqn-nond}, \eqref{density-eqn-nond}, \eqref{p_Vlasov-eqn-nond-gardner} and \eqref{n_Vlasov-eqn-nond-gardner} and equate different powers of $\epsilon$.  We temporarily drop the constant $\alpha_2$ in the subsequent expressions and equations. However, we note that the constants $\alpha_1,~\alpha_2$ and $\alpha_3$ will explicitly appear in the coefficients of the    Landau damping, nonlinear and the dispersive terms in the evolution equation.   The results will be given in the following subsections. 
\subsection{First-order perturbations and nonlinear wave speed}
\begin{figure}[ht]
\centering
{\includegraphics[height=2.0in,width=3.5in]{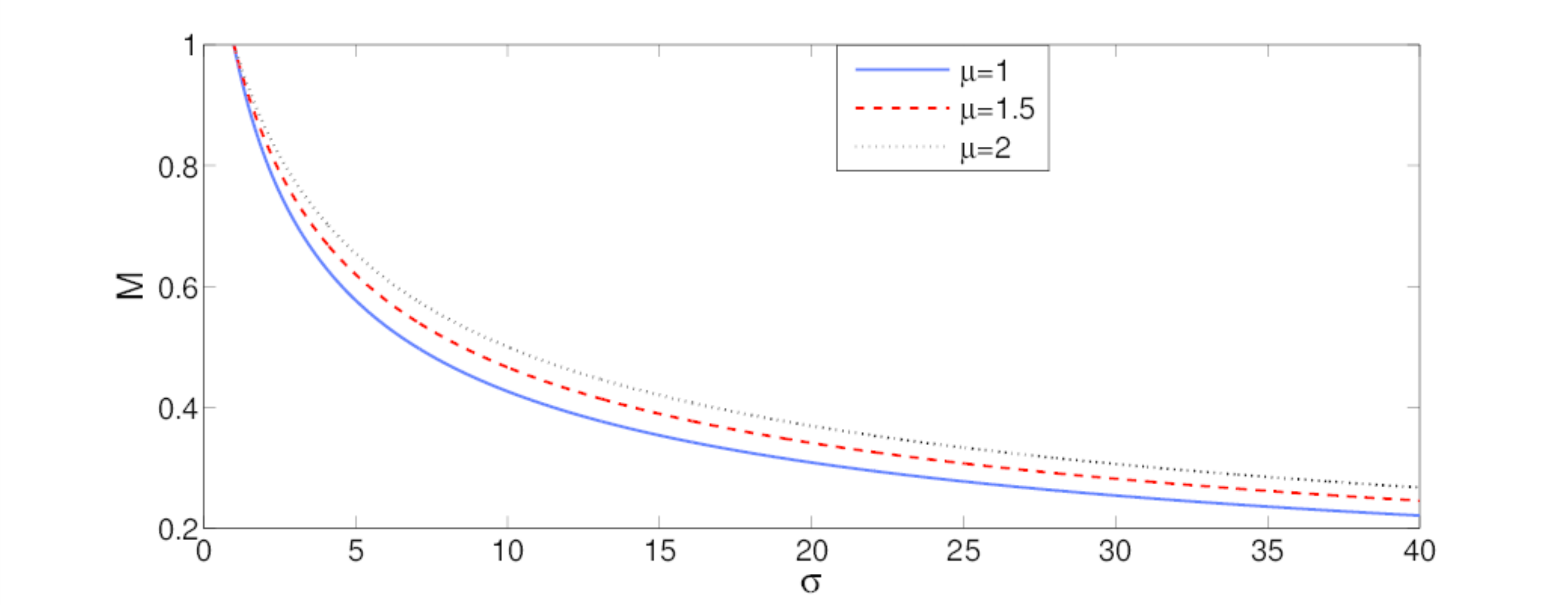}}
\caption{The linear phase velocity $M$ [Eq. \eqref{phase-velocity}] of  DAWs is shown with respect to the temperature ratio $\sigma~(=T_h/T_c)$ for different values of the density ratio $\mu~(=n_{h0}/n_{c0})$ as in the figure. }
\label{fig:fig2}
\end{figure}

We successively equate the coefficients of $\epsilon^2$ from Eqs. \eqref{cont-eqn-nond} and \eqref{montm-eqn-nond}, the coefficients of $\epsilon$ from Eqs. \eqref{Poisson-eqn-nond} and \eqref{density-eqn-nond},  and  the coefficients of $\epsilon^2$ from Eqs. \eqref{p_Vlasov-eqn-nond-gardner} and \eqref{n_Vlasov-eqn-nond-gardner}  to obtain
 \begin{eqnarray}
 n^{(1)}_d={u^{(1)}_d}/M,  \label{u^(1)-n^(1)} 
 \end{eqnarray}
 \begin{eqnarray}
 u^{(1)}_d=-{\phi^{(1)}}/{M}, \label{u^(1)-phi^(1)}
 \end{eqnarray}
   \begin{eqnarray}
 0=-(\mu_{cd}n^{(1)}_c+\mu_{hd}n^{(1)}_h)+n^{(1)}_d, \label{mu_nd-mu_bd}
 \end{eqnarray}
 \begin{eqnarray}
 n^{(1)}_j=\sqrt{m_jT_h/m_hT_j}\int_{-\infty}^{\infty}f^{(1)}_jdv,\label{n^(1)_p-n^(1)_n}
 \end{eqnarray}
   \begin{eqnarray}
 v\partial_{\xi} f^{(1)}_h+vf^{(0)}_h\partial_{\xi}\phi^{(1)}=0,\label{v_p-1}
 \end{eqnarray}
 \begin{eqnarray}
 v\partial_{\xi} f^{(1)}_c+\sigma vf^{(0)}_c\partial_{\xi}\phi^{(1)}=0.\label{v_n-1}
 \end{eqnarray}
 Equations \eqref{u^(1)-n^(1)} and \eqref{u^(1)-phi^(1)} reduce to
 \begin{eqnarray}
 n_d^{(1)}=-\phi^{(1)}/{M^2}.\label{nd1}
 \end{eqnarray}
 Also,    Eq. \eqref{v_p-1} yields the following relation \cite{ott1969} 
 \begin{eqnarray}
 \partial_\xi f^{(1)}_h=-f^{(0)}_h\partial_\xi\phi^{(1)}+\lambda(\xi,\tau)\delta(v),\label{for unique soln v_p}
 \end{eqnarray}
where $\delta(v)$ is the Dirac delta function and $\lambda(\xi, \tau)$ is an arbitrary function of $\xi$ and $\tau$ implying that the   solution for ${\partial f^{(1)}_h}/{\partial\xi}$   is not unique. Thus, for the unique solution to exist we  add a  higher-order term $\epsilon^6\alpha_1\left({\partial f^{(1)}_h}/{\partial\tau}\right)$ originating from the term $\epsilon^{5}\alpha_1\left({\partial f_h}/{\partial\tau}\right)$ in Eq. \eqref{p_Vlasov-eqn-nond-gardner} after the expressions  \eqref{stretch-gardner} and \eqref{expansions} being substituted. Thus, we rewrite Eq. \eqref{v_p-1} as  \cite{ott1969}
\begin{eqnarray}
\alpha_1\epsilon^2\partial_\tau f^{(1)}_{h\epsilon}+v\partial_\xi f^{(1)}_{h\epsilon}=-vf^{(0)}_h\partial_\xi\phi^{(1)},\label{v_p modfd-1}
\end{eqnarray}
Similarly, from Eq. \eqref{v_n-1}, we obtain
\begin{equation}
\alpha_1\epsilon^2\partial_\tau f^{(1)}_{c\epsilon}+v\partial_\xi f^{(1)}_{c\epsilon}=-\sigma vf^{(0)}_c\partial_\xi\phi^{(1)}.\label{v_n modfd-1}
\end{equation}
The unique solutions of Eqs. \eqref{v_p modfd-1} and \eqref{v_n modfd-1}  can be given   as
\begin{equation}
f^{(1)}_j=\lim_{\epsilon\rightarrow 0} f^{(1)}_{j\epsilon},\label{unique sol_1}
\end{equation}
where $f^{(1)}_{j\epsilon}$ $(j=h,~c)$ are to be obtained   by taking the Fourier transform of Eq. \eqref{v_p modfd-1} with respect to $\xi$ and $\tau$ according to the formula
\begin{equation}
\hat{f}(\omega, k)=\int^{\infty}_{-\infty}\int^{\infty}_{-\infty}f(\xi, \tau)e^{i(k\xi-\omega\tau)}d\xi d\tau. \label{FT}
\end{equation}
Thus, we obtain
\begin{equation}
\hat{f}^{(1)}_{h\epsilon}=-kvf^{(0)}_h\left(kv-\epsilon^2\alpha_1\omega\right)^{-1}\hat{\phi}^{(1)}. \label{1FT f_p}
\end{equation}
Next, to remove the singularity   in Eq. \eqref{1FT f_p}    we replace $\omega$ by $\omega+i\eta$ with  $\eta~(>0)$  being small, and obtain
\begin{equation}
\hat{f}^{(1)}_{h\epsilon}=-kvf^{(0)}_h\left[\left(kv-\epsilon^2\alpha_1\omega\right)-i\eta\alpha_1\epsilon^2\right]^{-1}\hat{\phi}^{(1)}. \label{2FT f_p}
\end{equation}
Proceeding to the limit as $\epsilon\rightarrow 0$,   using the Plemelj's formula (in which $P$   denotes the Cauchy principal value)
\begin{equation}
\lim_{\epsilon\rightarrow 0}\left(x+i\epsilon\right)^{-1}=-i\pi\delta(x)+\text{P} \left({1}/{x}\right),\label{Plmj formla}
\end{equation}
and noting that $x\text{P}(1/x)=1$, $x\delta(x)=0$, we obtain
\begin{equation}
\hat{f}^{(1)}_h=-f^{(0)}_h\hat{\phi}^{(1)}.\label{3FT f_p}
\end{equation}
Next, we take the Fourier inversion of Eq. \eqref{3FT f_p} to obtain
\begin{equation}
f^{(1)}_h=-f^{(0)}_h\phi^{(1)}.\label{4FT f_p}
\end{equation}
Similar expression for cold ions can also be obtained  from Eq. \eqref{v_n modfd-1}  as
\begin{equation}
f^{(1)}_c=-\sigma f^{(0)}_c\phi^{(1)}.\label{4FT f_n}
\end{equation}
Substituting the expressions from Eqs. \eqref{4FT f_p} and \eqref{4FT f_n}   into Eq. \eqref{n^(1)_p-n^(1)_n},  we  obtain
\begin{eqnarray}
n^{(1)}_h=-\phi^{(1)},\label{n^(1)_p}
\end{eqnarray}
\begin{eqnarray}
n^{(1)}_c=-\sigma\phi^{(1)}.\label{n^(1)_n}
\end{eqnarray}
The expressions for $n_j^{(1)}$ $(j=c,h,d)$ from Eqs. \eqref{nd1},  \eqref{n^(1)_p} and \eqref{n^(1)_n} can now be used in  Eq. \eqref{mu_nd-mu_bd}, to obtain the following expression for the nonlinear wave speed \cite{barman2014}
\begin{equation}
 M=\left(\mu_{hd}+\sigma\mu_{cd}\right)^{-1/2}\equiv\sqrt{(1+\mu)/(\mu+\sigma)}.\label{phase-velocity}\end{equation} 
Evidently,  $M$ has exactly the same expression   as $v_p$ obtained in the linear dispersion law \eqref{disp-relation} for plane wave perturbations. Figure \ref{fig:fig2} shows the characteristics of $M$ with $\sigma$ for different values of the density ratio $\mu$. We note that since   $\sigma\equiv T_h/T_c>1$, the quantity under the square root in Eq. \eqref{phase-velocity} is always smaller than unity,  i.e., the DAWs propagate with a phase speed being smaller than the DA speed $c_s$. We also find that  as the value of  $\mu$   increases, the value  of the phase velocity $M$  increases. However, the same decreases for increasing values of the temperature ratio $\sigma$, and approaches to zero for $\sigma\gg1$.  The latter is, however,   inadmissible in the present theory as it would not give any physical result.   We mention that the results in this subsection  (up to the linear theory) are the same as for the KdV and Gardner or mKdV equations. However, the results (nonlinear) will be different in the expressions for second and higher order perturbations.
\subsection{Second-order perturbations}
 We successively equate the coefficients of $\epsilon^{3}$ from Eqs. \eqref{cont-eqn-nond} and \eqref{montm-eqn-nond}, the coefficients of $\epsilon^2$ from Eqs. \eqref{Poisson-eqn-nond} and \eqref{density-eqn-nond}   as well as the coefficients of $\epsilon^{3}$ from Eqs. \eqref{p_Vlasov-eqn-nond} and \eqref{n_Vlasov-eqn-nond} to obtain
\begin{equation}
-Mn^{(2)}_d+u^{(2)}_d=-n^{(1)}_du^{(1)}_d, \label{M-u^(2)-n^(2)}
\end{equation}
\begin{equation}
u^{(2)}_d=-\left(1/M\right)\phi^{(2)}+\left(1/2M^3\right)\left(\phi^{(1)}\right)^2, \label{M-u^(2)-phi^(2)}
\end{equation}
\begin{equation}
n^{(2)}_d=\mu_{cd}n^{(2)}_c+\mu_{hd}n^{(2)}_h, \label{M-mu_nd-mu_bd}
\end{equation}
\begin{equation}
n^{(2)}_j=\sqrt{m_jT_h/m_hT_j}\int_{-\infty}^{\infty}f^{(2)}_jdv,\label{M-n^(2)_p-n^(2)_n}
\end{equation}
\begin{equation}
v\partial_{\xi} f^{(2)}_h-\partial_{\xi}\phi^{(1)}\partial_v f^{(1)}_h+vf^{(0)}_h\partial_{\xi}\phi^{(2)}=0,\label{M-v_p-1}
\end{equation}
\begin{equation}
v\partial_{\xi} f^{(2)}_c-(1/m)\partial_{\xi}\phi^{(1)}\partial_v f^{(1)}_c+\sigma vf^{(0)}_c \partial_{\xi}\phi^{(2)}=0.\label{M-v_n-1}
\end{equation}
Eliminating $u^{(2)}_d$ from Eqs. \eqref{M-u^(2)-n^(2)} and \eqref{M-u^(2)-phi^(2)} and   using the first-order expressions from  Eqs. \eqref{u^(1)-phi^(1)} and \eqref{nd1}, we obtain
\begin{equation}
n^{(2)}_d=(3/2M^4)\left(\phi^{(1)}\right)^2-(1/M^2)\phi^{(2)}.\label{M-n(2)_phi^(1)_phi^(2)}
\end{equation}
Substituting the expressions for $f_j$ from Eqs. \eqref{4FT f_p} and \eqref{4FT f_n} into Eqs. \eqref{M-v_p-1} and \eqref{M-v_n-1}  we successively obtain
\begin{equation}
v\partial_\xi f^{(2)}_h-vf^{(0)}_h \phi^{(1)}\partial_\xi\phi^{(1)}+vf^{(0)}_h\partial_\xi\phi^{(2)}=0,\label{M-v_p-2}
\end{equation}
\begin{equation}
v\partial_\xi f^{(2)}_c-\sigma^2vf^{(0)}_c \phi^{(1)}\partial_\xi\phi^{(1)}+\sigma vf^{(0)}_c\partial_\xi\phi^{(2)}=0.\label{M-v_n-2}
\end{equation}
As in the previous section, to obtain the unique solutions for $f_j^{(2)}$ we introduce  a higher-order term $\epsilon^{7}\alpha_1\left(\partial f^{(1)}_j/\partial\tau\right)$ originating from the term $\epsilon^{5}\alpha_1\left(\partial f_j/\partial\tau\right)$  in Eqs. \eqref{p_Vlasov-eqn-nond-gardner} and \eqref{n_Vlasov-eqn-nond-gardner} after the expressions  \eqref{stretch-gardner} and \eqref{expansions} being substituted, and replace $f^{(2)}_j$ by $f^{(2)}_{j\epsilon}$. Thus, we rewrite Eqs. \eqref{M-v_p-2} and \eqref{M-v_n-2} as
\begin{eqnarray}
&&\alpha_1\epsilon^2\partial_\tau f^{(2)}_{h\epsilon}+v\partial_\xi f^{(2)}_{h\epsilon}+vf^{(0)}_h\partial_\xi\phi^{(2)}\notag\\
&&=(1/2)vf^{(0)}_h\partial_\xi(\phi^{(1)})^2,\label{M-v_p-3}
\end{eqnarray}
\begin{eqnarray}
&&\alpha_1\epsilon^2\partial_\tau f^{(2)}_{c\epsilon}+v\partial_\xi f^{(2)}_{c\epsilon}+\sigma vf^{(0)}_c\partial_\xi\phi^{(2)}\notag\\
&&=(\sigma^2/2)vf^{(0)}_c\partial_\xi(\phi^{(1)})^2.\label{M-v_n-3}
\end{eqnarray}
As before, the solutions $f^{(2)}_j,~j=h,c$  can be found uniquely from   Eqs. \eqref{M-v_p-3} and \eqref{M-v_n-3}   by the following relation 
\begin{equation}
f^{(2)}_j=\lim_{\epsilon\rightarrow 0} f^{(2)}_{j\epsilon}.\label{M-unique sol_1}
\end{equation}
So, taking the Fourier transform of Eq. \eqref{M-v_p-3} according to the formula \eqref{FT},    and   replacing $\omega$ by $\omega+i\eta$ ($\eta>0$), we obtain
\begin{equation}
\hat{f}^{(2)}_{h\epsilon}=\left[-\hat{\phi}^{(2)}+\frac12(\hat{\phi}^{(1)})^2\right]\left[\frac{kvf^{(0)}_h}{(kv-\epsilon^2\alpha_1\omega)-i\eta\alpha_1\epsilon^2}\right]. \label{M-1FT f_p}
\end{equation}
Proceeding to the limit as $\epsilon\rightarrow 0$ and using the Plemelj's formula \eqref{Plmj formla} together with $x\text{P}(1/x)=1$ and $x\delta(x)=0$, we obtain from Eq. \eqref{M-1FT f_p} the following
\begin{equation}
\hat{f}^{(2)}_h=\left[-\hat{\phi}^{(2)}+(1/2)(\hat{\phi}^{(1)})^2\right]f^{(0)}_h.\label{M-2FT f_p}
\end{equation}
Next, the Fourier inverse transform of  Eq. \eqref{M-2FT f_p}  yields the following solution for the hot ion distribution function
\begin{equation}
f^{(2)}_h=\left[-\phi^{(2)}+(1/2)(\phi^{(1)})^2\right]f^{(0)}_h.\label{M-3FT f_p}
\end{equation}
By the same way as above, one can also obtain the similar expression for the cold ion distribution   as
\begin{equation}
f^{(2)}_c=\left[-\sigma\phi^{(2)}+(\sigma^2/2)\left(\phi^{(1)}\right)^2\right]f^{(0)}_c.\label{M-3FT f_n}
\end{equation}
The expressions for the second order perturbations of the number densities can now be obtained  from  Eq.      \eqref{M-n^(2)_p-n^(2)_n} by using Eqs. \eqref{M-3FT f_p} and \eqref{M-3FT f_n}  as
\begin{equation}
n^{(2)}_h=-\phi^{(2)}+(1/2)(\phi^{(1)})^2,\label{M-3FT n_p}
\end{equation}
and
\begin{equation}
n^{(2)}_c=-\sigma\phi^{(2)}+(\sigma^2/2)(\phi^{(1)})^2.\label{M-3FT n_n}
\end{equation}
Furthermore, substituting the expressions for $n^{(2)}_d$, $n^{(2)}_h$, $n^{(2)}_c$ from Eqs.   \eqref{M-n(2)_phi^(1)_phi^(2)}, \eqref{M-3FT n_p} and \eqref{M-3FT n_n} into Eq. \eqref{M-mu_nd-mu_bd}, we obtain for nonzero perturbations the following relation
\begin{eqnarray}
&&-(1/2)\left(\mu_{hd}+\mu_{cd}\sigma^2\right)(\phi^{(1)})^2+\left(\mu_{hd}+\mu_{cd}\sigma\right)\phi^{(2)}\notag\\
&&=(1/M^2)\phi^{(2)}-(3/2M^4)(\phi^{(1)})^2,\label{M-phi^(1)-phi^(2)}
\end{eqnarray}
in which the coefficient of $\phi^{(2)}$ vanishes by the dispersion relation \eqref{phase-velocity} yielding
\begin{equation}
\widetilde{b}\equiv(1/2)(\sigma\sigma_1)^2\left(3-\sigma_2/\sigma^2_1\right)\left(\phi^{(1)}\right)^2=0,\label{sigma_1-sigma_3} 
\end{equation}
where $\sigma_1=(\sigma+\mu)/\sigma(1+\mu)$ and $\sigma_2=(\sigma^2+\mu)/\sigma^2(1+\mu)$.
Here, we note that   since $\phi^{(1)}$ is of the order of $\epsilon$, the factor   $R_1\equiv3-\sigma_2/\sigma^2_1$ must be at least of the order of $\epsilon$, which we have already assumed in the beginning of this section. So,   the term $\widetilde{b}$  will contribute to the terms in the coefficient of $\epsilon^{3}$ of  the Poisson's equation in Sec. IVC [See Eq. \eqref{M2-mu_nd-mu_bd}], and finally in the coefficient $b_1$ of the Gardner equation \eqref{mK-dV}. 
\subsection{Third-order perturbations}
Equating the coefficients of $\epsilon^4$ from Eqs. \eqref{cont-eqn-nond} and \eqref{montm-eqn-nond}, the coefficients of $\epsilon^{3}$ from Eqs. \eqref{Poisson-eqn-nond} and \eqref{density-eqn-nond},  and  the coefficients of $\epsilon^4$ from Eqs. \eqref{p_Vlasov-eqn-nond-gardner} and \eqref{n_Vlasov-eqn-nond-gardner}, we successively obtain
\begin{eqnarray}
&&-M\partial_\xi n^{(3)}_d+\partial_\tau n^{(1)}_d+\partial_\xi u^{(3)}_d+\partial_\xi(n^{(1)}_du^{(2)}_d)\notag\\
&&+\partial_\xi(n^{(2)}_du^{(1)}_d)=0,\label{M-u^(3)-n^(3)}
\end{eqnarray}
\begin{equation}
\partial_\tau u^{(1)}_d-M\partial_\xi u^{(3)}_d+\partial_\xi(u^{(1)}_du^{(2)}_d)=\partial_\xi\phi^{(3)}.\label{M-u^(3)-phi^(3)}
\end{equation}
\begin{eqnarray}
\partial^2_\xi\phi^{(1)}=&&-(\mu_{hd}n^{(3)}_h+\mu_{cd}n^{(3)}_c)+ n^{(3)}_d\notag\\
&& +(1/2)(\sigma\sigma_1)^2\left(3-\sigma_2/\sigma^2_1\right)\left(\phi^{(1)}\right)^2,\label{M2-mu_nd-mu_bd}
\end{eqnarray}
\begin{equation}
n^{(3)}_j=\sqrt{m_jT_h/m_hT_j}\int_{-\infty}^{\infty}f^{(3)}_jdv.\label{M-n^(3)_p-n^(3)_n}
\end{equation}
\begin{eqnarray}
&&-\alpha_1M\partial_\xi f^{(1)}_h+v\partial_\xi f^{(3)}_h+vf^{(0)}_h\partial_\xi\phi^{(3)}\notag\\
&&-\partial_\xi\phi^{(2)}\partial_v f^{(1)}_h
-\partial_\xi\phi^{(1)}\partial_v f^{(2)}_h=0,\label{M-v_p-4}
\end{eqnarray}
\begin{eqnarray}
&&-\alpha_1M\partial_\xi f^{(1)}_c+v\partial_\xi f^{(3)}_c+\sigma vf^{(0)}_c\partial_\xi\phi^{(3)}\notag\\
&&-(1/m)\left(\partial_\xi\phi^{(2)}\partial_v f^{(1)}_c+\partial_\xi\phi^{(1)}\partial_v f^{(2)}_c\right)=0.\label{M-v_n-4}
\end{eqnarray}
As before, to get the unique solutions for $f^{(3)}_j$ (for $j=h,~c$), we introduce a   higher-order term $\epsilon^8\alpha_1\left(\partial f^{(3)}_j/\partial\tau\right)$ originating from the term $\epsilon^{5}\alpha_1\left(\partial f_j/\partial\tau\right)$ in Eqs. \eqref{p_Vlasov-eqn-nond-gardner} and \eqref{n_Vlasov-eqn-nond-gardner}, and then replacing $f^{(3)}_j$ by $f^{(3)}_{j\epsilon}$, where
\begin{equation}
f^{(3)}_j=\lim_{\epsilon\rightarrow 0} f^{(3)}_{j\epsilon},\label{unique sol_3}
\end{equation}
we obtain from Eqs. \eqref{M-v_p-4} and \eqref{M-v_n-4} as
\begin{eqnarray}
&&\alpha_1\epsilon^2\partial_\tau f^{(3)}_{h\epsilon}+v\partial_\xi f^{(3)}_{h\epsilon}+vf^{(0)}_h\partial_\xi\phi^{(3)}\notag\\
&&=(D_{ha}+vD_{hb})f^{(0)}_h,\label{M-v_p-5}
\end{eqnarray}
\begin{eqnarray}
&&\alpha_1\epsilon^2\partial_\tau f^{(3)}_{c\epsilon}+v\partial_\xi f^{(3)}_{c\epsilon}+\sigma vf^{(0)}_c\partial_\xi\phi^{(3)}\notag\\
&&=(D_{ca}+vD_{cb})f^{(0)}_c,\label{M-v_n-5}
\end{eqnarray}
where we have used   Eqs.  \eqref{4FT f_p} and \eqref{M-3FT f_p} to obtain Eq. \eqref{M-v_p-5}, and Eqs.                   \eqref{4FT f_n} and \eqref{M-3FT f_n} to obtain  Eq. \eqref{M-v_n-5}. The expressions for $D_{ha}$, $D_{hb}$, $D_{ca}$ and $D_{cb}$ are given by
\begin{eqnarray}
&&D_{ha}=-\alpha_1M\partial_\xi\phi^{(1)},\notag\\
&&D_{hb}=-(1/6)\partial_\xi(\phi^{(1)})^3+\partial_\xi\left(\phi^{(1)}\phi^{(2)}\right),\label{M-D_j(a,b)}\\
&&D_{ca}=-\alpha_1M\sigma\partial_\xi\phi^{(1)},\notag\\
&&D_{cb}=-(\sigma^3/6)\partial_\xi(\phi^{(1)})^3+\sigma^2\partial_\xi\left(\phi^{(1)}\phi^{(2)}\right).\notag
\end{eqnarray}
Next, taking the  Fourier transforms of  Eq. \eqref{M-v_p-5} with respect to $\xi$ and $\tau$ according to the formula \eqref{FT} and   replacing $\omega$ by $\omega+i\eta$ ($\eta>0$), we obtain
\begin{eqnarray}
\hat{f}^{(3)}_{h\epsilon}=-\left[\frac{kvf^{(0)}_h\hat{\phi}^{(3)}+i(\hat{D}_{ha}+v\hat{D}_{hb})f^{(0)}_h}{(kv-\epsilon^2\alpha_1\omega)-i\eta\alpha_1\epsilon^2}\right]. \label{M-4FT f_p}
\end{eqnarray}
In the limit of  $\epsilon\rightarrow 0$,   Eq. \eqref{M-4FT f_p} yields
\begin{eqnarray}
&&\hat{f}^{(3)}_h+f^{(0)}_h\hat{\phi}^{(3)}=-i\left[\text{P}\left(1/kv\right)+i\pi(\text{sgn}~k/k)\delta(v)\right]\times\notag\\
&&(\hat{D}_{ha}+v\hat{D}_{hb})f^{(0)}_h,\label{M-5FT f_p}
\end{eqnarray}
where we have used the  Plemelj's formula \eqref{Plmj formla} together with $x\text{P}(1/x)=1$, $x\delta(x)=0$ and $\delta(kv)=\left(\text{sgn}~k/k\right)\delta(v)$.
Now, multiplying both sides of Eq. \eqref{M-5FT f_p} by $ik$ and  then integrating with respect to $v$, we obtain
\begin{equation}
ik\left(\hat{n}^{(3)}_h+\hat{\phi}^{(3)}\right)=\hat{D}_{hb}+i\sqrt{\pi/2}~\text{sgn}(k)\hat{D}_{ha},\label{M-n^(3)_p}
\end{equation}
which under the Fourier inverse transform  reduces  to
\begin{eqnarray}
\partial_\xi n^{(3)}_h+\partial_\xi\phi^{(3)}=&&\partial_\xi(\phi^{(1)}\phi^{(2)})-(1/6)\partial_\xi(\phi^{(1)})^3\notag\\
&&+\sqrt{\pi/2}~F^{-1}\left(i~\text{sgn}(k)\hat{D}_{ha}\right).\label{M-n^(3)_p-phi^(3)}
\end{eqnarray}
Here,  $F^{-1}$ stands for the Fourier inverse transform. Using the convolution theorem of Fourier transform and noting that $F^{-1}[i~\text{sgn}(k)]=-(1/\pi)\text{P}\left(1/\xi\right)$, we obtain from Eq. \eqref{M-n^(3)_p-phi^(3)}
\begin{eqnarray}
\partial_\xi n^{(3)}_h&&+\partial_\xi\phi^{(3)}=\partial_\xi(\phi^{(1)}\phi^{(2)})-(1/6)\partial_\xi(\phi^{(1)})^3\notag\\
&&+\alpha_1M\left(1/\sqrt{2\pi}\right)P\int^{\infty}_{-\infty}\partial_{\xi'}\phi^{(1)}\frac{d\xi'}{\xi-\xi'}.\label{M-n^(3)_p-phi^(3)-P}
\end{eqnarray}
Similar equation for the third-order perturbations can be obtained for cold ions as  
\begin{eqnarray}
&&\left(1/\sqrt{m\sigma}\right)\partial_\xi n^{(3)}_c+\sqrt{\frac{\sigma}{m}}\partial_\xi\phi^{(3)}=\frac{1}{\sqrt{m\sigma}}\left[\sigma^2\partial_\xi\left(\phi^{(1)}\phi^{(2)}\right)\right.\notag \\
&&\left.-\frac{\sigma^3}{6}\partial_\xi\left(\phi^{(1)}\right)^3\right]+\frac{\alpha_1M\sigma}{\sqrt{2\pi}}
P\int^{\infty}_{-\infty}\partial_{\xi'}\phi^{(1)}\frac{d\xi'}{\xi-\xi'}.\label{M-n^(3)_n-phi^(3)-P}
\end{eqnarray}
\subsection{Gardner and mKdV equations  with Landau damping}
In order to obtain the evolution (Gardner) equation for first-order perturbations, we first eliminate $u^{(3)}_d$ and $n^{(3)}_d$   from Eqs. \eqref{M-u^(3)-n^(3)}-\eqref{M2-mu_nd-mu_bd} to obtain 
\begin{eqnarray}
&&\partial^3_{\xi}\phi^{(1)}=-\mu_{hd}\partial_{\xi} n^{(3)}_h-\mu_{cd}\partial_{\xi} n^{(3)}_c-(1/M^2)\partial_{\xi}\phi^{(3)}\notag\\
&&+\left(1/M\right)\left[\partial_{\tau} n^{(1)}_d+\left(1/M\right)\left(\partial_\tau u^{(1)}_d+\partial_\xi\left(u^{(1)}_du^{(2)}_d\right)\right)\right.\notag\\
&&\left.+\partial_\xi\left(n^{(1)}_du^{(2)}_d+n^{(2)}_du^{(1)}_d\right)\right]\notag\\
&&+(\sigma\sigma_1)^2\left(3-{\sigma_2}/{\sigma^2_1}\right)\phi^{(1)}\partial_\xi\phi^{(1)}.\label{M-n^(3)_j}
\end{eqnarray}
Next, we use   Eqs. \eqref{M-n^(3)_p-phi^(3)-P} and \eqref{M-n^(3)_n-phi^(3)-P} to eliminate $n^{(3)}_h$ and $n^{(3)}_c$ from Eq. \eqref{M-n^(3)_j}, and   obtain
\begin{eqnarray}
&&\partial^3_\xi\phi^{(1)}=-\left(\mu_{hd}+\sigma^2\mu_{cd}\right)\partial_\xi\left(\phi^{(1)}\phi^{(2)}\right)\notag\\
&&+(1/6)\left(\mu_{hd}+\sigma^3\mu_{cd}\right)\partial_\xi\left(\phi^{(1)}\right)^3\notag\\
&&-\frac{\alpha_1M}{\sqrt{2\pi}}\left(\mu_{hd}+\sqrt{m}\sigma^{3/2}\mu_{cd}\right)
P\int^{\infty}_{-\infty}\partial_{\xi'}\phi^{(1)}\frac{d\xi'}{\xi-\xi'}\notag\\
&&+(1/M)\left[\partial_\tau n^{(1)}_d+(1/M)\left(\partial_\tau u^{(1)}_d+\partial_\xi\left(u^{(1)}_du^{(2)}_d\right)\right)\right.\notag\\
&&\left.+\partial_\xi\left(n^{(1)}_du^{(2)}_d+n^{(2)}_du^{(1)}_d\right)\right]\notag\\
&&+(\sigma\sigma_1)^2\left(3-\sigma_2/\sigma^2_1\right)\phi^{(1)}\partial_\xi\phi^{(1)}.\label{M-n^(3)_j2}
\end{eqnarray}
 Next,  we substitute the expressions of $u_d^{(2)}$ and $n_d^{(2)}$  from Eqs. \eqref{M-u^(2)-phi^(2)} and \eqref{M-n(2)_phi^(1)_phi^(2)} into Eq. \eqref{M-n^(3)_j2}. Here, we note that the coefficient of $\phi^{(2)}\phi^{(1)}$ is $\sim \left(3-\sigma_2/\sigma_1^2\right)\sim\epsilon$, and so the term $\propto\left(3-\sigma_2/\sigma_1^2\right)\left(\phi^{(2)}\phi^{(1)}\right)\sim\epsilon^4$ [since $\phi^{(2)}\sim\epsilon^2$ and $\phi^{(1)}\sim\epsilon$ ]. Thus, this term will not contribute to Eq. \eqref{M-n^(3)_j2} and we discard it. Finally, we replace $\phi^{(1)}$ by $n\equiv n_d^{(1)}$ using   Eqs. \eqref{u^(1)-n^(1)} and \eqref{u^(1)-phi^(1)}  to obtain  from Eq. \eqref{M-n^(3)_j2}  the following Gardner (KdV-mKdV) equation     
\begin{eqnarray}
\frac{\partial n}{\partial\tau}&&+a\text{P}\int^{\infty}_{-\infty}\frac{\partial n}{\partial\xi'}\frac{d\xi'}{\xi-\xi'}\notag\\
&&+b_1n\frac{\partial n}{\partial\xi}+b_2n^2\frac{\partial n}{\partial\xi}+c\frac{\partial^3 n}{\partial\xi^3}=0,\label{mK-dV}
\end{eqnarray}
where  the coefficients $a$ and $c$, corresponding to the  Landau damping and dispersive effects, are the same as for the KdV equation \eqref{K-dV}. The  nonlinear coefficients $b_1$ and $b_2$ are, however, different and given by 
\begin{equation}
b_1=\left(3-\sigma_2/\sigma_1^2\right)/2\sqrt{\sigma\sigma_1},\label{b1}
\end{equation}
\begin{equation}
b_2=(15/4)\left(\alpha_2/(\sigma\sigma_1)^{1/2}\right)\left[1-\left(\sigma_3/15\sigma_1^3\right)\right].\label{b2}
\end{equation}
Here, the expressions for $\sigma_1$ and $\sigma_2$ are given before and
$\sigma_3=(\sigma^3+\mu)/\sigma^3(1+\mu)$.  
If we neglect the strength of the Landau damping i.e., if we set $\alpha_1=0$,   the term $\propto a$ in Eq. \eqref{mK-dV} vanishes, and   Eq. \eqref{mK-dV} reduces to the usual KdV-mKdV or Gardner equation for the propagation of DASWs in dusty bi-ion plasmas.
Equation \eqref{mK-dV} is the required   evolution equation for the first-order dust  density perturbations when the plasma parameters $\sigma$ and $\mu$ assume values close to the critical values, i.e.,   $R_1\sim o(\epsilon)$. In fact, this equation is  well applicable to DASWs when the plasma parameters assume the critical values or   are very close to or around the critical values of $\sigma$ and $\mu$.    
 
On the other hand, when   $R_1=0$, i.e., when the parameters $\sigma$ and $\mu$ assume the critical values along the thin (or black) curve as in Fig. \ref{fig:fig1}, the evolution of DASWs will no longer be describable by the KdV equation \eqref{K-dV}, but the mKdV equation, which can be obtained by  setting $b_1=0$ in Eq. \eqref{mK-dV}   as 
\begin{equation}
\frac{\partial n}{\partial\tau}+a\text{P}\int^{\infty}_{-\infty}\frac{\partial n}{\partial\xi'}\frac{d\xi'}{\xi-\xi'}+b_2n^2\frac{\partial n}{\partial\xi} 
+c\frac{\partial^3 n}{\partial\xi^3}=0,\label{mkdv}
\end{equation}
where the coefficients $a,~b_2$ and $c$ are the same as in the Gardner equation \eqref{mK-dV}.
We remind here that the evolution of DASWs with the effects of Landau damping are governed by the  KdV, mKdV or Gardner equations  depending on whether the density ratio $\mu$ and the temperature ratio $\sigma$  assume their values well below the critical values, very close to the critical values or near, around and exactly the critical values of $\mu$ and  $\sigma$  along the curve in Fig. \ref{fig:fig1}. Note that the mKdV equation is obtained on the assumption that $R_1=0$. However, when $R_2\equiv 1-\sigma_3/15\sigma_1^3=0$, i.e., the nonlinear term $b_2$ in the mKdV equation vanishes    in the $\sigma-\mu$ plane (See the thick or red curve in Fig. \ref{fig:fig1}), then the mKdV equation  fails to describe the evolution of DASWs. From Fig. \ref{fig:fig1}, it is also clear that there are some values of $\sigma$ and $\mu$ which may be closer to both the curves (e.g.,  the lower parts of the curves for $R_1=0$ and $R_2=0$ below the  point of intersection). In these parameter regimes both the KdV and mKdV equations  may not  be appropriate for the description of DASWs in plasmas.  Thus, the Gardner equation, being the generalized form of KdV and mKdV, describes the evolution of DASWs in dusty bi-ion plasmas for a wide range of values of $\sigma$ and $\mu$ including the critical values as in Fig. \ref{fig:fig1}.    
  \section{Landau damping rate}
\begin{figure}[ht]
\centering
\includegraphics[height=2in,width=3.5in]{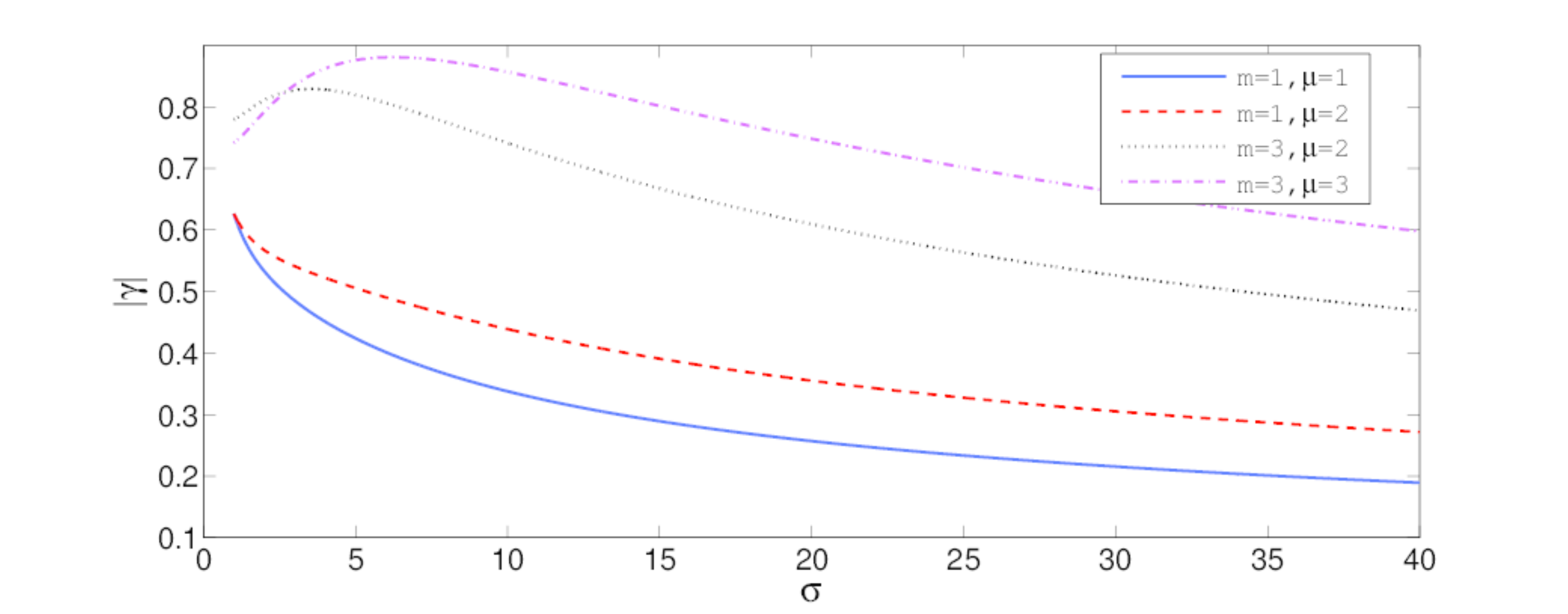}
\caption{The linear Landau damping rate $|\gamma|=\pi a$ is plotted against the temperature ratio $\sigma$ for different values of the mass and the density ratios $m$ and $\mu$ as in the figure.}
\label{fig:fig3}
\end{figure}

To obtain the Landau damping rate for DASWs we take the Fourier transform of Eq. \eqref{mK-dV} with $b_1,~b_2$ and $c$ being set to zero, according to the formula \eqref{FT}, and use the result that the inverse transform of $i~\text{sgn}(k)$ is $-(1/\pi)~\text{P}(1/\xi)$. Thus, we  obtain the Landau damping rate as \cite{barman2014} $|\gamma|=\pi a$. This damping rate is shown in Fig. \ref{fig:fig3} with respect to the temperature ratio $\sigma$, and for different values of the density ratio  $\mu$ and the mass ratio $m$. We find that for dusty plasmas with equal mass of  positive ions, the value  of $|\gamma|$ decreases (increases) with an increasing value  of the parameter $\sigma$ $(\mu)$. However, the case with unequal mass of ions $(m>1)$ is quite distinctive. In this case,   the damping rate initially increases as long as $1<\sigma<5$, and for $\sigma>5$, the behaviors  of $|\gamma|$ remain similar to the case of equal mass of ions.    Thus, in contrast to dusty pair-ion plasmas \cite{barman2014} or dusty electron-ion plasmas, an enhancement of  the hot ion concentration  (or a reduction of the  number of cold ions into the dust grain surface  in order to maintain the charge neutrality), increases the Landau damping rate in dusty bi-ion plasmas. 
\section{Solitons with Landau damping}
We first obtain an approximate analytic soliton solution of the Gardner equation \eqref{mK-dV} with a small effect of the Landau damping. Here, we assume that     $1\sim b_1,b_2\sim c\gg a\gg\epsilon$. This can easily be verified by considering   dusty plasma parameters satisfying $\sigma,~\mu>1$. Now, in absence of the Landau damping  Eq. \eqref{mK-dV}  reduces to
\begin{equation}
\partial_{\tau'} n+b_1'n\partial_\xi n+b_2'n^2\partial_\xi n+\partial^3_\xi n=0,\label{Grdnr eq.2}
\end{equation}
where we have rescaled the time variable $\tau$ and the coefficients  as $\tau'=\tau c$, $b_1'=b_1/c$, $b_2'=b_2/c$.
A  traveling wave solution  of Eq. \eqref{Grdnr eq.2}  can be obtained as \cite{wazwaz2007}
\begin{equation}
n(\xi, \tau')=6\xbar{U}\left[b_1'\pm\xbar{N}\text{cosh}\left((\xi-\xbar{U}\tau')/\xbar{W}\right)\right]^{-1},\label{sol of Gardner eq}
\end{equation}
where $\xbar{W}=\sqrt{6b_2'/(\xbar{N}^2-b_1'^2})$ is the constant width and $\xbar{U}=(\xbar{N}^2-b_1'^2)/6b_2'$ is the constant phase speed  (normalized by $c_s$) of the DASWs.
\begin{figure*}[ht]
\subfigure[]{\includegraphics[height=2in,width=3.4in]{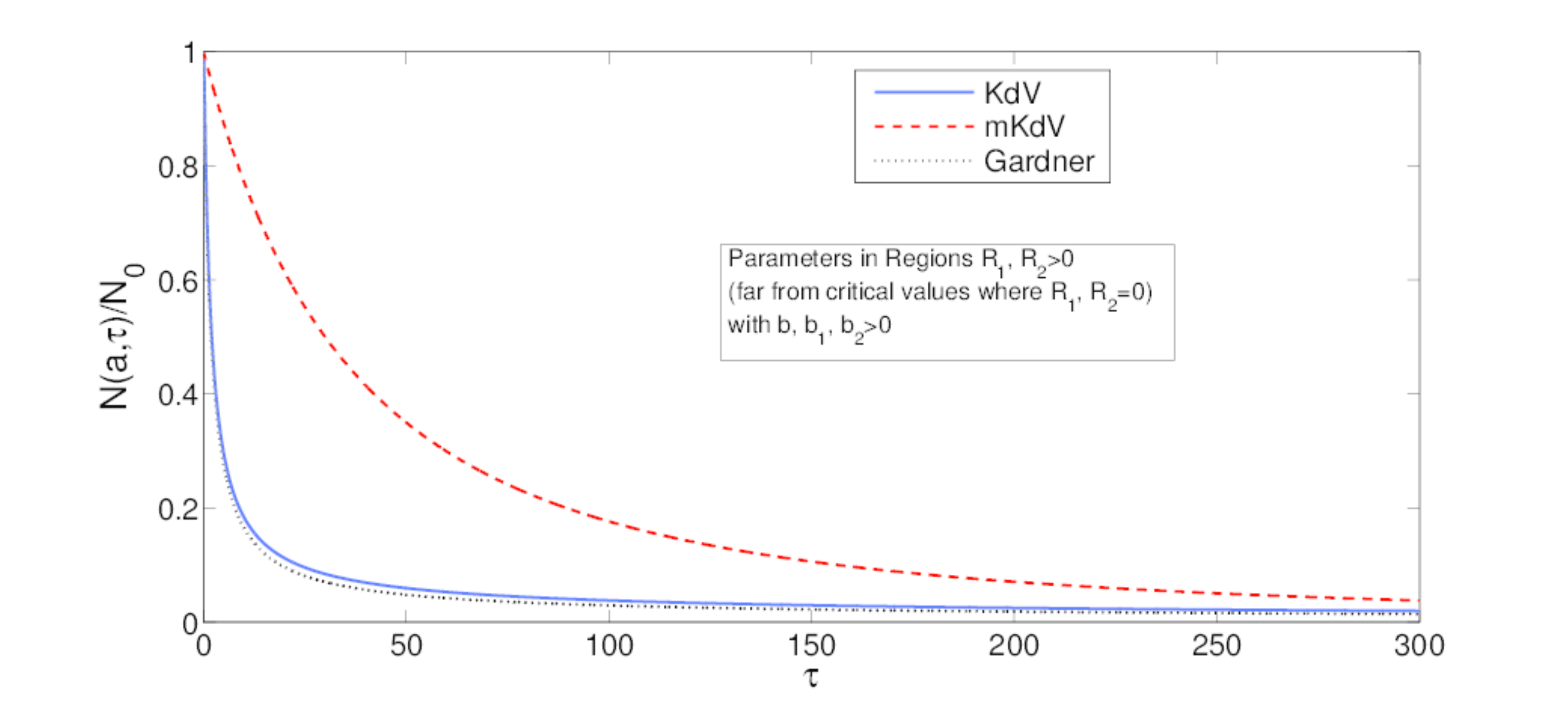}
\label{fig:fig4a}}  
  \subfigure[]{\includegraphics[height=2in,width=3.4in]{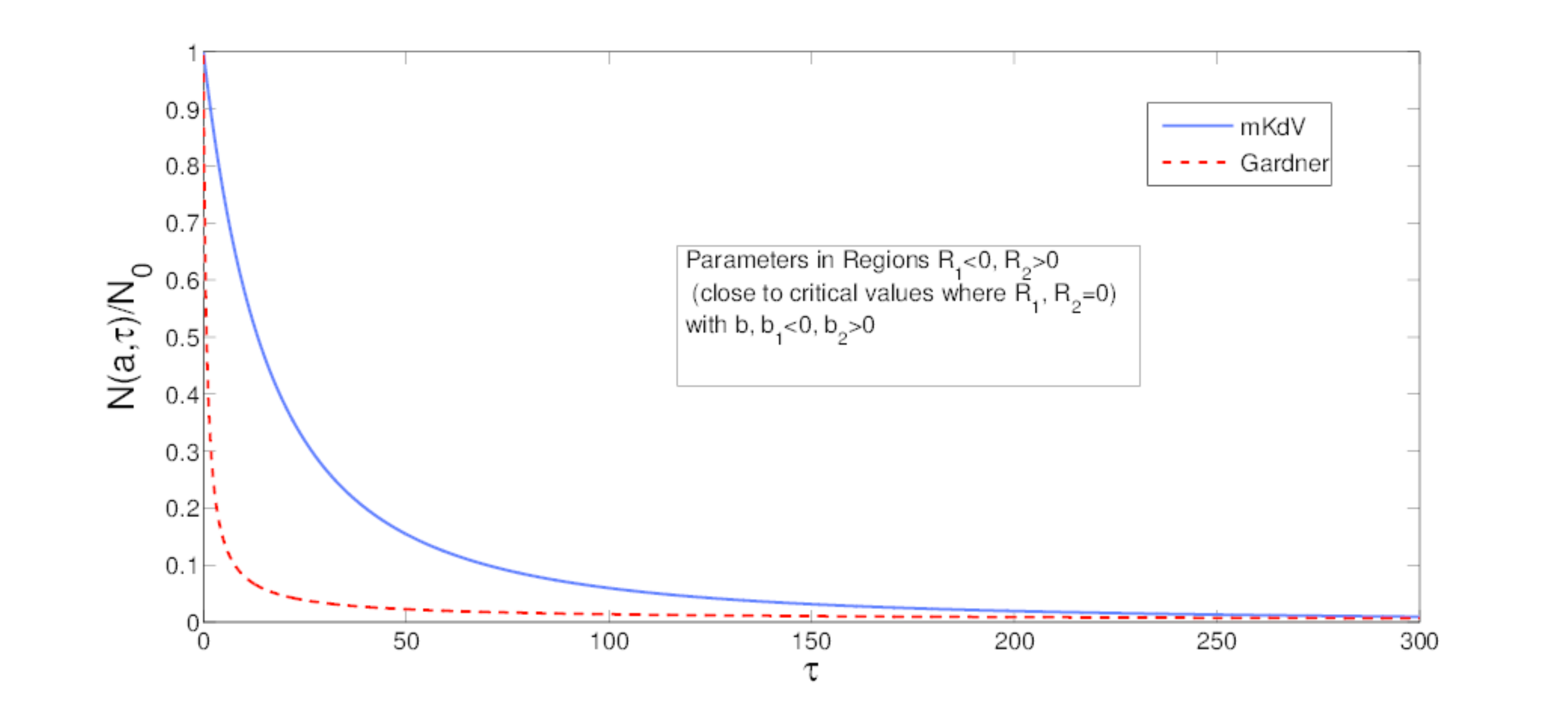}
\label{fig:fig4b}} 
\subfigure[]{\includegraphics[height=2in,width=3.4in]{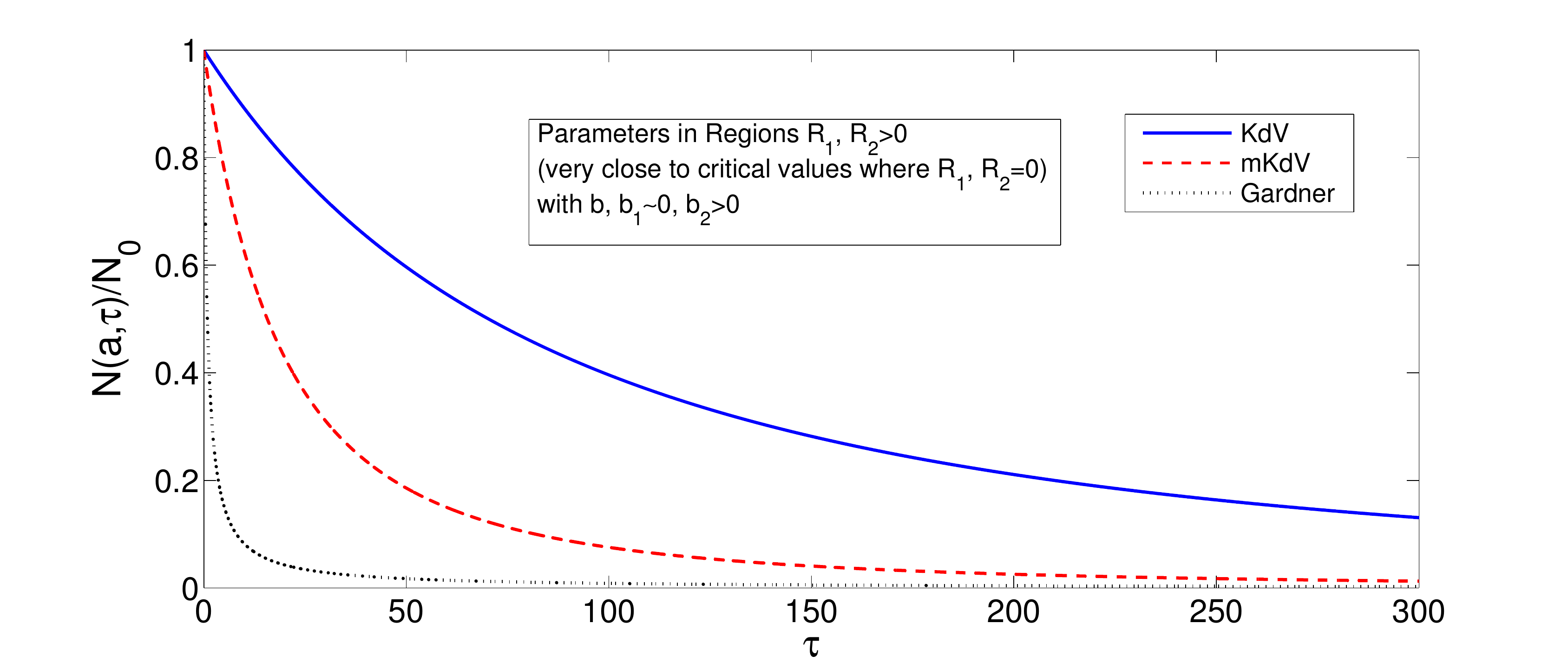}
\label{fig:fig4c}}
\caption{The  solitary wave amplitudes $N(a,\tau)$ (normalized to their equilibrium values $N_0$) with the effect of Landau damping are shown for KdV, mKdV and Gardner solitons [Eqs. \eqref{amp-kdv}, \eqref{amp-mkdv}, \eqref{M-solution2}]. Subfigure \ref{fig:fig4a} shows that when $\mu$ and $\sigma$ assume values $(m=\sigma=3,~\mu=7.4)$ well below the critical values on the curve $R_1=0$ and fall in the region $R_1>0$, the decay rates of the amplitudes of the  KdV and Gardner solitons with respect to $\tau$ are approximately the same as expected. However, the rate for the mKdV soliton is quite higher than the KdV and Gardner solitons. For values of $\mu$ and $\sigma$ $(m=3,~\sigma=15,~\mu=4.5)$ close to the critical values, but in the regions $R_1,~R_2<0$ for which $b,~b_1<0$ (and hence KdV equation fails to govern the dynamics of dust-acoustic solitons), subfigure \ref{fig:fig4b} shows that the gap between the decay rates of the Gardner and mKdV solitons becomes lower than that in Fig. \ref{fig:fig4a}. On the other hand,  when $\mu$ or $\sigma$ values $(m=3,~\sigma=11,~\mu=5.3)$ are very close to the critical values, but still in the regions $R_1,~R_2>0$ for which      $b,~b_1\approx0$ and $b_2>0$, subfigure \ref{fig:fig4c} shows that the gaps among the amplitudes of the KdV, mKdV and Gardner solitons again increase, and Gardner equation seems to be more appropriate for the description of dust-acoustic solitons. }
 \label{fig:fig4}
 \end{figure*}
It can easily be verified that Eq. \eqref{mK-dV} conserves the total number of particles  and an initial perturbation of the form \eqref{sol of Gardner eq}  decays to zero (See for some details, e.g., Ref. \onlinecite{barman2014}). The latter implies that  the wave amplitude $\xbar{N}$ is not a constant but decreases slowly with time. Thus,  we introduce a new space coordinate in a frame moving with the solitary wave and normalized to its width as \cite{barman2014}
\begin{equation}
z=\sqrt{(\xbar{N}^2-b_1'^2)/6b_2'}\left(\xi+\frac{b_1'^2}{3b_2'}\tau'-\frac{1}{6b_2'}\int_{0}^{\tau'}\xbar{N}^2d\tau'\right),\label{M-space coordinate}
\end{equation}
where $\xbar{N}=\xbar{N}(a,~\tau')$ is assumed to vary slowly with time and   $n=n(z,~\tau')$. Under this transformation   Eq. \eqref{mK-dV} reduces to
\begin{eqnarray}
&&\partial_{\tau'} n+\left[-A+Bz\left(d\xbar{N}/d\tau'\right)\right]\partial_z n+\left(b_1'/\xbar{W}\right)n\partial_z n\notag\\
&&+\left(b_2'/\xbar{W}\right)n^2\partial_z n+\left(1/\xbar{W}^3\right)\partial^3_z n\notag\\
&&+\left(a'/\xbar{W}\right)P\int_{-\infty}^{\infty}\partial_z' n\frac{dz'}{z-z'}=0,\label{M-after subst SP}
\end{eqnarray}
where $A=(\xbar{N}^2-b_1'^2)/6b_2'\xbar{W}$, $B=\xbar{N}/(\xbar{N}^2-b_1'^2)$, $a'=a/c$ and we have used $\partial n/\partial z'=\partial n/\partial z$ at $z=z'$.
 
In what follows, we investigate the solution of Eq. \eqref{M-after subst SP}, and introduce a multiple time scale analysis with respect to $a'$.  Thus, we write \cite{barman2014,bandyopadhyay2002}
\begin{equation}
n(z, \tau')=n^{(0)}+a'n^{(1)}+a'^2n^{(2)}+a'^3n^{(3)}+\cdots,\label{M-multiple time scale}
\end{equation}
where $n^{(i)},~i=0, 1, 2, 3,\cdots$ are  functions of $\tau'=\tau_0, \tau_1, \tau_2, \tau_3,\cdots$ in which 
$\tau_i=a'^{i}\tau'$. So, we have $\partial n^{(0)}/\partial\tau'=(\partial n^{(0)}/\partial\tau_0)(\partial \tau_0/\partial\tau')+(\partial n^{(0)}/\partial\tau_1)(\partial \tau_1/\partial\tau')+\cdots$, $\partial n^{(1)}/\partial\tau'=(\partial n^{(1)}/\partial\tau_0)(\partial \tau_0/\partial\tau')+(\partial n^{(1)}/\partial\tau_1)(\partial \tau_1/\partial\tau')+\cdots$ etc. and the similar expression   for $\partial\xbar{N}/\partial\tau$.
Substituting the expansion  \eqref{M-multiple time scale} into Eq. \eqref{M-after subst SP} and using the expressions for $\partial n^{(0)}/\partial\tau'$, $\partial n^{(1)}/\partial\tau'$ and $\partial \xbar{N}/\partial\tau'$  we obtain
\begin{eqnarray}
&&\left(\partial_{\tau'} n^{(0)}+a'\partial_{\tau_1} n^{(0)}+\cdots\right)+a'\left(\partial_{\tau'} n^{(1)}+a'\partial_{\tau_1} n^{(1)}+\cdots\right)\notag\\
&&+\left[-A+Bz\left(\partial_{\tau'}\bar{N}+a'\partial_{\tau_1}\bar{N}+\cdots\right)\right]\times\notag\\
&&\partial_z\left(n^{(0)}+a'n^{(1)}+\cdots\right)+\left(1/\xbar{W}\right)\left(b_1'+b_2'n^{(0)}\right)n^{(0)}\partial_z n^{(0)}\notag\\
&&+\frac{a'}{\xbar{W}}\left[b_1'\partial_z\left(n^{(1)}n^{(0)}\right)+b_2'n^{(0)}\left(n^{(0)}\partial_z n^{(1)}+2n^{(1)}\partial_z n^{(0)}\right)\right]\notag\\
&&+\left(1/\xbar{W}^3\right)\partial^3_z n^{(0)}+a'\left(1/\xbar{W}^3\right)\partial^3_z n^{(1)}\notag\\
&&+a'\frac{1}{\xbar{W}}P\int_{-\infty}^{\infty}\partial_z' n^{(0)}\frac{dz'}{z-z'}+\cdots=0,\label{M-big equation}
\end{eqnarray}
Equating the coefficients of the zeroth and first-order  of $a'$, we successively obtain   from Eq. \eqref{M-big equation} as
\begin{equation}
\beta\left[\partial_{\tau'}+(Bz)\partial_{\tau'}\xbar{N}\partial_z\right]n^{(0)}+M_0\partial_z n^{(0)}=0,\label{M-0th order eq}
\end{equation}
\begin{equation}
\beta\left[\partial_{\tau'}+(Bz)\partial_{\tau'}\xbar{N}\partial_z\right]n^{(1)}+\partial_z M_0n^{(1)}=\beta Rn^{(0)},\label{M-1st order eq}
\end{equation}
where  $\beta=\xbar{W}^3$ and
\begin{equation}
M_0=\partial^2_z+\left[n^{(0)}\left(b_1'+b_2'n^{(0)}\right)\left(A\xbar{W}\right)^{-1}-1\right],\label{M-bar}
\end{equation}
\begin{eqnarray}
 R n^{(0)}=-\left[\partial_{\tau_1} n^{(0)}+(Bz)\partial_{\tau_1}\xbar{N}\partial_z n^{(0)}\right.\notag\\
\left.+\frac{1}{\xbar{W}}P\int_{-\infty}^{\infty}\partial_z' n^{(0)}\frac{dz'}{z-z'}\right].\label{R-bar-n^(0)}
\end{eqnarray}\\
Using the  boundary conditions, namely, $n^{(0)}$, $\partial n^{(0)}/\partial z$, $\partial^2 n^{(0)}/\partial z^2$, $\partial^3 n^{(0)}/\partial z^3\longrightarrow 0$ as $z\rightarrow\pm\infty$ it can be shown that  $n^{(0)}=6U_0/(b_1'\pm\xbar{N}\cosh z)$ is a soliton solution of the equation $M_0\left(\partial n^{(0)}/\partial z\right)=0$. Thus,  $n^{(0)}=6U_0/(b_1'\pm\xbar{N}\cosh z)$ will be a soliton solution of Eq. \eqref{M-0th order eq} if and only if 
${\partial\xbar{N}}/{\partial\tau}=0$. So,  Eq. \eqref{M-1st order eq} reduces to 
\begin{equation}
\beta\left[\partial_\tau n^{(1)}\right]+\partial_z M_0 n^{(1)}=\beta Rn^{(0)}.\label{M-1st order eq-AC}
\end{equation}\\
Next, for the existence of a solution of Eq. \eqref{M-1st order eq-AC}, $Rn^{(0)}$  must be orthogonal to all solutions $g(z)$  of $L^{+}[g]=0$ which satisfy $g(\pm\infty)=0$, where $L^{+}$ is the operator adjoint to $L\equiv(\partial/\partial z)\bar{M}$, defined by 
\begin{equation}
\int_{-\infty}^{\infty}\psi_1(z)L[\psi_2(z)]dz=\int_{-\infty}^{\infty}\psi_2(z)L^{+}[\psi_1(z)]dz,\label{M-adjt oprtr}
\end{equation}
where $\psi_1(\pm\infty)=\psi_2(\pm\infty)=0$,    the operator $L^{+}$ is given by 
\begin{equation}
L^{+}\equiv\frac{\partial^3}{\partial z^3}+\left[\frac{6\xbar{N}(\xbar{N}\pm b_1'\cosh z)}{(b_1'\pm\xbar{N}\cosh z)^2}-1\right]\frac{\partial}{\partial z}, \label{adjoint operator}
\end{equation}
and the only solution of $L^{+}[g]=0$ with $g(\pm\infty)=0$ is $g(z)=(\xbar{N}^2-b_1'^2)/\xbar{N}(b_1'\pm\xbar{N}\cosh z)$. Thus, we obtain
\begin{equation}
\int_{-\infty}^{\infty}\frac{1}{\xbar{N}}(\xbar{N}^2-b_1'^2)(b_1'\pm\xbar{N}\cosh z)^{-1}Rn^{(0)}dz=0.\label{M-eqn}
\end{equation}
Substituting the expression for $Rn^{(0)}$ from Eq. \eqref{R-bar-n^(0)} into Eq. \eqref{M-eqn}, we get
\begin{widetext} 
\begin{eqnarray}
&&\pm\frac{\partial\xbar{N}}{\partial\tau_1}\int_{-\infty}^{\infty}\left[\frac{(\xbar{N}^2-b_1'^2)}{\bar{N}^2}\frac{\xbar{N}\text{cosh}~z}{(b_1'\pm\xbar{N}\text{cosh}~z)^3}
+\frac{\xbar{N}z~\text{sinh}~z}{(b_1'\pm\xbar{N}\text{cosh}~z)^3}\right]dz\notag\\
&&-\frac{(\xbar{N}^2-b_1'^2)}{\xbar{N}~\xbar{W}}
\text{P}\int_{-\infty}^{\infty}\int_{-\infty}^{\infty}\frac{1}{(b_1'\pm\xbar{N}\cosh z)}\frac{\partial}{\partial z'}\left[\frac{1}{(b_1'\pm\xbar{N}\cosh z')}\right]\frac{dzdz'}{z-z'}=0,\label{rev1}
\end{eqnarray}
\end{widetext}
from which    we obtain after few steps the following 
\begin{widetext}
\begin{eqnarray}
&&\frac{\partial \xbar{N}}{\partial\tau_1}\left[\frac{(3\xbar{N}^2-2b_1'^2)}{2\xbar{N}^2}\int_{-\infty}^{\infty}\frac{dz}{(b_1'\pm\xbar{N}\text{cosh}~z)^2}
-b_1'\frac{(\xbar{N}^2-b_1'^2)}{\xbar{N}^2}\int_{-\infty}^{\infty}\frac{dz}{(b_1'\pm\xbar{N}\text{cosh}~z)^3}\right]dz\notag\\
&&-\frac{(\xbar{N}^2-b_1'^2)^{3/2}}{\xbar{N}\sqrt{6b_2'}}
\text{P}\int_{-\infty}^{\infty}\int_{-\infty}^{\infty}\frac{1}{(b_1'\pm\xbar{N}\cosh z)}\frac{\partial}{\partial z'}\left[\frac{1}{(b_1'\pm\xbar{N}\cosh z')}\right]\frac{dzdz'}{z-z'}=0.\label{rev2}
\end{eqnarray}
\end{widetext}
Now, we perform the integrations as follows 
\begin{widetext}
\begin{eqnarray}
\int_{-\infty}^{\infty}(b_1'\pm\xbar{N}\text{cosh}~z)^{-2}dz=\left[2/(\xbar{N}^2-b_1'^2)^{3/2}\right]
\left[2b_1'\text{tan}^{-1}\left((b_1'\mp\xbar{N})/\sqrt{\xbar{N}^2-b_1'^2}\right)+\sqrt{\xbar{N}^2-b_1'^2}\right],\label{rev3}
\end{eqnarray}
\begin{eqnarray}
\int_{-\infty}^{\infty}(b_1'\pm\xbar{N}\text{cosh}~z)^{-3}dz=\left[2/(\xbar{N}^2-b_1'^2)^{5/2}\right]
\left[(2b_1'+\xbar{N}^2)\text{tan}^{-1}\left((b_1'\mp\xbar{N})/\sqrt{\xbar{N}^2-b_1'^2}\right)+\frac{3b_1'}{2}\right].\label{rev4}
\end{eqnarray}
\end{widetext}
Finally, using the results in Eqs.~\eqref{rev3} and \eqref{rev4}, we obtain from Eq.~\eqref{rev2} as
\begin{eqnarray}
&&\frac{\partial\xbar{N}}{\partial \tau_1}\lambda_1-\lambda_2\text{P}\int_{-\infty}^{\infty}\int_{-\infty}^{\infty}\frac{1}{(b_1'\pm\xbar{N}\cosh z)}\times\notag\\
&&\partial_z'\left(b_1'\pm\xbar{N}\cosh z'\right)^{-1}\left[dz'/(z-z')\right]dz=0,\label{M-1st ordr diffnal eq}
\end{eqnarray}
where $\lambda_1$ and $\lambda_2$ are given by
\begin{eqnarray}
&&\lambda_1=\left[1/(\xbar{N}^2-b_1'^2)^{3/2}\right]\left[8b_1'\text{tan}^{-1}\left((b_1'\mp\xbar{N})/\sqrt{\xbar{N}^2-b_1'^2}\right)\right.\notag\\
&&\left.+\left((3\xbar{N}^2+b_1'^2)/\xbar{N}^2\right)\sqrt{\xbar{N}^2-b_1'^2}\right], \label{lambda1}
\end{eqnarray}
\begin{equation}
\lambda_2=\left[(\xbar{N}^2-b_1'^2)^{3/2}/\xbar{N}\sqrt{6b_2'}\right].\label{lambda2}
\end{equation}
Equation \eqref{M-1st ordr diffnal eq} is a first-order (in $\tau$) partial differential equation for the solitary wave amplitude $\xbar{N}(a, \tau)$ whose exact analytic solution is much   complicated. However, one can consider some particular cases as follows:  

We note that for small-amplitude perturbations $\epsilon|n_1(\xi,\tau)|\ll1$. However, inspecting on the solution \eqref{sol of Gardner eq} of Eq. \eqref{mK-dV} in absence of the Landau damping effect, we find that this condition may be satisfied when either $\xbar{U}$ is   small  or the denominator of the expression on the right-hand side becomes larger. Also, for real values of $\xbar{W}$ and $\xbar{U}~(>0)$, we have $\xbar{N}^2>b_1'^2$ when $\xbar{N}$ is independent on time $\tau$. However, when $\xbar{N}\equiv\xbar{N}(a,\tau)$, we must have  $\xbar{N}^2(a,\tau)>b_1'^2$ $\forall\tau$ for $\lambda_1$ and $\lambda_2$  to be real. Here, we consider the  case in which $\xbar{N}^2(a,\tau)\gg b_1'^2$ $\forall\tau$.    In this case, though some generalization may be lost, the basic qualitative features \cite{barman2014} of the  solution of Eq. \eqref{M-1st ordr diffnal eq} will remain the same.     Under this approximation  the solution of   Eq. \eqref{M-1st ordr diffnal eq} is given by a quadratic equation in $\xbar{N}$ which yields two solutions (for $\zeta=\pm1$) 
\begin{equation}
\xbar{N}(a, \tau)=(3b_2/c)^{1/4}\xbar{N}_0\left(\lambda_3/\lambda_4\right),\label{M-solution2}
\end{equation}
where $\xbar{N}=\xbar{N}_0$ at $\tau=0$ and 
\begin{eqnarray}
&&\lambda_3=3^{5/4}(b_2/c)^{1/4}\xbar{N}_0+\zeta\left[(3b_2/c)^{1/2}\right.\notag\\
&&\left.\times(3\xbar{N}_0\mp\pi b_1/c)^2\pm2\sqrt{2}\pi b_1 A'a\xbar{N}^2_0\tau/c\right]^{1/2},\label{lambda3}
\end{eqnarray}
\begin{equation}
\lambda_4=\sqrt{3b_2/c}\left(6\xbar{N}_0\mp2\pi b_1/c\right)-\sqrt{2} A'a\xbar{N}^2_0\tau, \label{lambda4}
\end{equation}
 \begin{equation}
A'=\text{P}\int_{-\infty}^{\infty}\int_{-\infty}^{\infty}\text{sech}z\frac{\partial}{\partial z'}(\text{sech} z')\frac{dz'}{z-z'}dz.\label{A'}
\end{equation}  
We mention that the upper and lower signs in the expressions for $\lambda_1,~\lambda_3$ and $\lambda_4$ are corresponding to $\pm$ in the Gardner solution \eqref{sol of Gardner eq}. In our numerical investigation, we will, however, consider $\zeta=1$ and  the lower (upper) sign for the existence of solitons in different regions as in  Fig. \ref{fig:fig4a} (Figs. \ref{fig:fig4b} and \ref{fig:fig4c}) to be shown shortly.   Clearly,  a decrement of the wave amplitude of the Gardner soliton is  seen to occur.  

On the other hand, in absence of the Landau damping effects (i.e., when $a=0$), the KdV   and mKdV equations  [Eqs. \eqref{K-dV} and \eqref{mkdv}]   have the following traveling wave solutions 
\begin{equation}
n=N~\text{sech}^2\left[\left(\xi-U\tau\right)/W\right],\label{sol-kdv}
\end{equation}
\begin{equation}
n=\widetilde{N}~\text{sech}\left[\left(\xi-\widetilde{U}\tau\right)/\widetilde{W}\right],\label{sol-mkdv}
\end{equation}
where $N=3U/b$ and $\widetilde{N}=\sqrt{6\widetilde{U}/b'_2}$ are the constant amplitudes, $W=\sqrt{12c/Nb}\equiv\sqrt{4c/U}$ and $\widetilde{W}=\sqrt{1/\widetilde{U}}$ are the constant widths, and $U$ and $\widetilde{U}$ are the constant phase speeds (normalized by $c_s$) of the KdV and mKdV solitons.

Following the same procedure as above and assuming that  is a small parameter and $1\sim b,b_2\sim c\gg a\gg\epsilon$,   analytic solutions of Eqs. \eqref{K-dV} and \eqref{mkdv} can also be obtained. Thus, for the KdV equation \eqref{K-dV} we have the following solution \cite{barman2014}
\begin{equation}
n=N(a,\tau)~\text{sech}^2\left[\left(\xi-\frac{b}{3}\int_0^{\tau}Nd\tau\right)/W\right]+o(a),\label{sol-final-kdv}
\end{equation}
where 
\begin{equation}
N(a,\tau)=N_0\left[1+(\tau/\tau_0)\right]^{-2}, \label{amp-kdv}
\end{equation} 
with $N=N_0$ at $\tau=0$ and $\tau_0$ being given by
\begin{equation}
\tau_0^{-1}=\frac{a}{4}\sqrt{\frac{bN_0}{3c}}~\text{P}\int^{\infty}_{-\infty}\int^{\infty}_{-\infty}\frac{\text{sech}^2z}{z-z'}\frac{\partial}{\partial z'}\left(\text{sech}^2z'\right)dzdz'. \label{tau0-kdv}
\end{equation}
Also, for the mKdV equation \eqref{mkdv},  a substitution of $b_1=0$ in Eqs. \eqref{M-1st ordr diffnal eq}-\eqref{lambda2} gives rise the following solution 
\begin{equation}
n=\widetilde{N}(a,\tau)~\text{sech}\left[\left(\xi-\frac{1}{6b'_2}\int_0^{\tau'}\widetilde{N}^2d\tau'\right)/\widetilde{W}\right]+o(a),\label{sol-final-mkdv}
\end{equation}
where 
\begin{equation}
\widetilde{N}(a,\tau)=\widetilde{N}_0\left[1-(\tau/\widetilde{\tau})\right]^{-2}, \label{amp-mkdv}
\end{equation} 
with $\widetilde{N}=\widetilde{N}_0$ at $\tau=0$ and $\widetilde{\tau}$ being given by
\begin{equation}
\widetilde{\tau}^{-1}=\frac{a\widetilde{N}_0}{3\sqrt{6b'_2}}~\text{P}\int^{\infty}_{-\infty}\int^{\infty}_{-\infty}\frac{\text{sech}z}{z-z'}\frac{\partial}{\partial z'}\left(\text{sech}z'\right)dzdz'. \label{tau0-mkdv}
\end{equation}

We numerically investigate the time variations of the wave amplitudes   of the KdV, mKdV and Gardner solitons given by Eqs. \eqref{sol-final-kdv}, \eqref{sol-final-mkdv} and \eqref{M-solution2} respectively. The results are shown in Fig. \ref{fig:fig4} for  different   parameter regimes as applicable for the existence of KdV, mKdV and Gardner solitons. We note that when the values of $\sigma$ and $\mu$ fall in the common regions of $R_1,~R_2>0$ (for which $b,~b_1$ and $b_2>0$), but well below the critical values as in Fig. \ref{fig:fig1}, the KdV equation suffices to describe the nonlinear evolution of DA solitons. This is evident from Fig. \ref{fig:fig4a}.  Here, it is seen that    the behaviors of the  decay rates of the KdV and Gardner  solitons are   almost similar having their values close to each other (See the solid and dotted lines), however, the rate becomes  higher (in magnitude) for the mKdV solitons. Also, the KdV soliton points to lower wave amplitude than the mKdV and Gardner solitons at large $\tau$. Thus, in the  parameter regimes with $\sigma$ and $\mu$ far below the critical values,   mKdV  equation may not be relevant, however, the KdV and Gardner equations  can give rise to approximately the same results for the evolution of  DA solitons in plasmas.      When values of $\sigma$ and $\mu$ are close to the critical values but in the regions $R_1<0,~R_2>0$ (for which $b,~b_1<0$ and $b_2>0$), the KdV solitons cease to exist (because $b$ must be positive for   $\tau_0$  to be real). In this case, equations involving higher-order nonlinearities, namely Gardner or mKdV equation is necessary for the evolution of DA solitons.     Figure \ref{fig:fig4b} shows that the Gardner equation is more appropriate than the mKdV equation  as it  gives rise to solitons with lower amplitudes than the mKdV solitons.  On the other hand, when $\sigma$ and $\mu$ assume values very close to the critical values   but still in the common regions of $R_1,~R_2>0$ (for which $b,~b_1\approx0$ and $b_2>0$), Fig. \ref{fig:fig4c} shows that the decay rates of the KdV and mKdV solitons  get  higher (in magnitude) than that of the     Gardner soliton. In this case,    KdV equation fails, however, Gardner equation gives better (than the mKdV equation) results for the properties of DA solitons with Landau damping.      It is also found that (not shown in the figure) as the mass ratio $m$ increases, the amplitudes $N(a,\tau),~\widetilde{N}(a,\tau)$ and $\xbar{N}(a,\tau)$ for KdV, mKdV and Gardner solitons decrease. Also, as  $\sigma$ increases, both $N(a,\tau)$ and $\xbar{N}(a,\tau)$ increase, but   $\widetilde{N}(a,\tau)$ decreases. Furthermore,     an enhancement of $N(a,\tau)$ and a decrement of $\widetilde{N}(a,\tau)$ and $\xbar{N}(a,\tau)$ are seen to occur with an increase of the density ratio $\mu$. 

\section{Conclusion}
We have investigated the nonlinear propagation of small-amplitude dust-acoustic solitary waves in a dusty plasma with the effects of Landau damping associated with the finite inertial effects of hot and cold ions. By considering the lower and higher order perturbations, the evolution of dust-acoustic solitons are shown to be governed by KdV, mKdV or Gardner (KdV-mKdV) equation. The existence of the critical values of $\sigma$ and $\mu$ (denoting, respectively, the temperature and the density ratios of hot to cold ions)   in the $\sigma-\mu$ parameter plane for which the KdV and mKdV solitons cease to exist is given.   It is found that the KdV solitons with Landau damping exist only for values of  $\sigma$ and $\mu$  well below their critical values. Beyond the KdV limit, i.e., when $\sigma$ and $\mu$ assumes values close to their critical values, the Gardner equation is seen to be more appropriate than the mKdV equation for the evolution of small-amplitude dust-acoustic solitons in bi-ion dusty plasmas. It is also found that regardless of the ion masses equal or not, there exist a common value, i.e.,  $\sigma,~\mu\approx10$ for which both the KdV and mKdV equations fail to govern the dust-acoustic solitons in plasmas. The properties of the  wave phase velocity and the Landau damping rate  are also studied with different values of  $\sigma$, $\mu$ and the mass ratio $m$ of cold to hot ions. It is shown that the damping rate is quite different for dusty plasmas with equal and unequal masses of positive ions.  Distinctive features of the KdV, mKdV and Gardner solitons with a small (but finite) effect of the Landau damping are compared for appropriate parameter values of $\sigma$, $\mu$ and  $m$. It is seen that beyond the KdV limit, Gardner equation always points to smaller wave amplitudes with the effect of Landau damping. 
The results should be useful for understanding the collisionless wave damping of dust-acoustic solitons in laboratory as well as space plasmas such as those in planetary rings (e.g., the F-ring of Saturn) where negatively charged dusts and two species of positive ions are the major constituents.
\acknowledgments{A.B. thanks University Grants Commission (UGC), Government of India, for Rajib Gandhi National Fellowship (RGNF) with Ref. No. F1-17.1/2012-13/RGNF-2012-13-SC-WES-17295/(SA-III/Website). } 

\end{document}